\documentclass[11pt]{article}
\begin{document}
\vspace*{-.6in}
\thispagestyle{empty}
\begin{flushright}
hep-th/0601101
\end{flushright}
\baselineskip = 18pt

\vspace{.5in} {\Large
\begin{center}
\textbf{Conserved Charges in Even Dimensional Asymptotically locally Anti-de Sitter Space-times}
\end{center}}

\begin{center}
Rong-Gen Cai$^{\star,}$\footnote{Email address: cairg@itp.ac.cn}
and Li-Ming Cao$^{\star,\dagger,}$\footnote{Email address:
caolm@itp.ac.cn}
\end{center}

\begin{center}
\emph{$^{\star}$ Institute of Theoretical Physics, Chinese Academy
of Sciences, \\ P.O. Box 2735, Beijing 100080, China \\
$^{\dagger}$ Graduate School of the Chinese Academy of Sciences,
Beijing 100039, China}

\vspace{.5in}


\end{center}

\vspace{.2in}

\begin{center}
\underline{ABSTRACT}
\end{center}
\begin{quotation}
\noindent Based on the recent paper hep-th/0503045, we derive a
formula of calculating conserved charges in even dimensional
asymptotically {\it locally} anti-de Sitter space-times by using
the definition of Wald and Zoupas. This formula generalizes the
one proposed by Ashtekar {\it et al}. Using the new formula we
compute the masses of Taub-Bolt-AdS space-times by treating
Taub-Nut-AdS space-times as the reference solution. Our result
agrees with those resulting from ``background subtraction" method
or ``boundary counterterm" method. We  also calculate the
conserved charges of Kerr-Taub-Nut-AdS solutions in four
dimensions and higher dimensional Kerr-AdS solutions with Nut
charges. The mass of (un)wrapped brane solutions in any dimension
is given.
\end{quotation}

\vfil
\centerline{}

\newpage

\addtocontents{toc}{\protect\setcounter{tocdepth}{2}}
\pagenumbering{arabic}

\vspace{.5in}
\section{Introduction}
Asymptotically AdS space-times have been investigated thoroughly
in recent years due to the AdS/CFT correspondence~\cite{AdS, GKP,
Witten}. It relates the gravity in asymptotic AdS space-times to
dual conformal field theories (CFTs) living on the boundary of the
 AdS space-times. In this correspondence, the boundary fields
which set the boundary conditions of bulk fields are identified
with CFT sources which couple to gauge invariant operators. For
example, the boundary metric plays the role of the metric of the
space-time on which the dual field theory is defined; on the other
hand, it is a source of the energy momentum tensor of boundary
CFT. This requires the existence of space-times associated with
general Dirichlet boundary conditions for the metric. Such general
boundary conditions include the so-called asymptotically {\it
locally} anti-de Sitter (AlAdS) space-times, which require only
asymptotically {\it locally} AdS, not exact AdS. Asymptotically
exact AdS case has been studied in the literatures, for example
see \cite{ht,am}, and hereafter we call them AAdS space-times for
simplicity. Note that for the $n$-dimensional AAdS space-times,
the boundary has the topology ${\bf R}\times {\bf S}^{n-2}$, while
for the AlAdS space-times, the topology of their boundary needs
not to be ${\bf R}\times {\bf S}^{n-2}$. Due to the difference
between the boundary topologies, some methods, which are very
powerful to calculate conserved charges for AAdS space-times, do
not work for AlAdS space-times.

Therefore, it is interesting to study the conserved charges in
AlAdS space-times in its own right.  In the literatures, there are
different methods to obtain conserved charges for AAdS
space-times, see for example, references~\cite{ht,am,Abbott,HH,
by, bk, AD,Olea}. The comparison among these notions of conserved
charges  in AAdS space-times has been made by Hollands {\it et
al.} in a recent paper~\cite{Hollands}. For AAdS space-times, the
method of Ashtekar {\it et al} and the method of ``boundary
counterterm" are independent of the reference background. Using
these two methods, one therefore needs not consider this reference
background problem and can get the conserved charges of AAdS
space-times by straightforward calculations. However, for  the
AlAdS space-times, those methods which rigorously depend upon the
boundary conditions of AAdS space-times would invalidate, because
the boundary conditions of AlAdS space-times may be very different
from those of AAdS space-times.  For example, the method of
Ashtekar {\it et al.} \cite{am, AD} does not work for AlAdS
space-times because the boundary topology $\mathbf{R}\times
\mathbf{S}^{n-2}$ is required in this method.  Moreover, the
definition of $n$-dimensional AAdS space-times requires a
condition that the product of  the conformal factor to the power
of $(3-n)$ and the Weyl tensor of the unphysical space-time admits
a smooth limit on the conformal boundary~\cite{AD}. This condition
can not be fulfilled for general AlAdS space-times. For example.
one can easily check that the method of Ashtekar {\it et al.} does
not work for the asymptotically AdS space-times with Nut charges,
which are typical AlAdS space-times and will be discussed in the
present paper.

Several methods have been proposed by some authors to define the
conserved charges in AlAdS space-times, such as the ``holographic
charges" studied in~\cite{PS3}, the method given in~\cite{Aros,
Aros1}, and the superpotential method~\cite{Deru}.  In this paper,
based on the recent work of Hollands {\it et al}~\cite{Hollands},
we develop a method of calculating conserved charges in even
dimensional AlAdS space-times by using the covariant phase space
definition of Wald and Zoupas~\cite{Zwald}. This method
generalizes the formula of Ashtekar {\it et al.} such that we can
calculate conserved charges in even dimensional AlAdS space-times.
This method is background dependent, and one has to specify a
reference background before calculating the conserved charges for
these AlAdS space-times.

This paper is organized as follows. In the next section, we
briefly review the definition of conserved charges given by Wald
and Zoupas~\cite{Zwald}, and give explicit forms of some related
quantities in AlAdS space-times. In Sec.~3, we derive the formula
of calculating conserved charges in even dimensional AlAdS
space-times by using the analysis of Hollands {\it et
al.}~\cite{Hollands}. Our formula generalizes the one given by
Ashtekar {\it et al.}. In Sec.~4, using this new formula, we
calculate the masses of Taub-Bolt-AdS solutions by treating the
Taub-Nut-AdS space-times as the reference background. In Sec.~5,
four dimensional Kerr-Taub-Nut-AdS solution is discussed, and the
mass and angular momentum associated to it are calculated. Sec.~6
is devoted to calculating the conserved charges for higher
dimensional Kerr-AdS solutions with Nut charges. In Sec.~7 we give
the mass for (un)wrapped black brane solutions. We end in Sec.~8
with conclusion and discussion.

\section{Wald's Definition and the Boundary of AlAdS Space-times}

In differential covariant theories of gravity, Wald {\it et
al.}\cite{Zwald} developed a general prescription to define
``conserved charges" at asymptotic boundaries for any space-times.
In this paper, we will use this method to define the conserved
charges dual to some asymptotic symmetry generators of AlAdS
space-times. For simplicity, we consider no matter case (i.e.,
$g_{ab}$ are the only dynamical fields), therefore the
corresponding differential covariant Lagrangian of $n$-dimensional
AlAdS space-times $(M,g_{ab})$ is:
\begin{equation}
\mathbf{L}=\frac{1}{16\pi G}\left[R-2\Lambda\right]
\mbox{{\boldmath $\epsilon$}},
\end{equation}
where we have put the Lagrangian in the form of differential form
and $\mbox{{\boldmath $\epsilon$}}$ is the volume element. The
variation of the Lagrange density $\mathbf{L}$ can be written as
\begin{equation} \delta\mathbf{L}=\mathbf{E}^{ab}\delta
g_{ab}+d \mathbf{\Theta},
\end{equation}
where $\mathbf{\Theta}$ is an $(n-1)$-form, which is called {\it
symplectic potential form}, and it is a local linear function of
field variation. $\mathbf{E}^{ab}$ corresponds to the equations of
motion.  Their explicit forms are
\begin{equation}
\mathbf{\Theta}_{a_1\cdots a_{n-1}}(g_{ab},\delta
g_{ab})=\frac{1}{16 \pi G}\mbox{{\boldmath $\epsilon$}}_{a_1\cdots
a_{n-1}a} \mbox{{\boldmath $v$}}^{a}(g_{ab},\delta g_{ab}),
\end{equation}
\begin{equation} \mathbf{E}^{ab}=\frac{1}{16 \pi G}
\left[R^{ab}-\frac{1}{2}R g^{ab}+\Lambda
g^{ab}\right]\mbox{{\boldmath $\epsilon$}},
\end{equation}
where
\begin{equation}
\mbox{{\boldmath $v$}}^{a}=g^{ab} \nabla ^c \delta g_{cb} -
g^{bc}\nabla^a \delta g_{bc}.
\end{equation}
The {\it symplectic current $(n-1)-$form} $\mbox{{\boldmath
$\omega$}}$ is defined by taking an antisymmetrized variation of
$\mathbf{\Theta}$:
\begin{equation}
\mbox{{\boldmath $\omega$}}_{a_1\cdots
a_{n-1}}(g_{ab},\delta_{1}g_{ab}, \delta_{2}g_{ab}) =\frac{1}{16
\pi G}\mbox{{\boldmath $\epsilon$}}_{a_1\cdots a_{n-1}a}
\mbox{{\boldmath $w$}}^{a} (g_{ab},\delta_{1}g_{ab},
\delta_{2}g_{ab}),
\end{equation}
where $\mbox{{\boldmath $w$}}$ is a 1-form given by
\begin{equation}
\mbox{{\boldmath $w$}}^{a} (g_{ab},\delta_{1}g_{ab},
\delta_{2}g_{ab}) =
P^{abcdef}(\delta_{1}g_{bc}\nabla_{d}\delta_{2}g_{ef}-\delta_{2}g_{bc}\nabla_{d}\delta_{1}g_{ef}),
\end{equation}
where
\begin{equation}
P^{abcdef}=g^{ae} g^{fb} g^{cd}-\frac{1}{2} g^{ad} g^{be}
g^{fc}-\frac{1}{2} g^{ab} g^{cd} g^{ef}-\frac{1}{2} g^{bc} g^{ae}
g^{fd}+\frac{1}{2} g^{bc} g^{ad} g^{ef}.
\end{equation}
The integral of the symplectic current form over an
$(n-1)-$dimensional submanifold $\Sigma$ on $\widetilde{M}$ gives
the {\it presymplectic form},
\begin{equation}
\Omega_{\Sigma}(g, \delta_1 g, \delta_2 g) = \int_{\Sigma}
\mbox{{\boldmath $\omega$}}(g, \delta_1 g, \delta_2 g),
\end{equation}
where $\widetilde{M}=M \cup \mathcal{B}$ is the conformal
completion of $M$ with the boundary manifold $\mathcal{B}$. The
presymplectic structure $\Omega_{\Sigma}$ does not depend on the
choice of  $\Sigma$ if $\delta_1 g$ and $\delta_2 g$ satisfy
linearized field equations and $g$ has suitable asymptotic
condition~\cite{Zwald,Lwald,Iwald,Iwald95}. Here we have assumed
that ``kinetically" allowed field space $\mathcal{F}$ has been
defined such that $\mbox{{\boldmath $\omega$}}$ can be extended
continuously to $\mathcal{B}$ for all $\delta_1 g$ and $\delta_2
g$ tangent to the solution subspace $\overline{\mathcal{F}}$ and
$\Sigma$ has an unambiguous boundary $\partial\Sigma$ in
$\mathcal{B}$.

The ``conserved charges" $H_\xi:\overline{\mathcal{F}} \rightarrow
\mathbf{R}$ associated with a vector field $\xi^a$ representing an
asymptotic symmetry defined by using presymplectic form
in~\cite{Zwald} satisfies
\begin{equation}
\label{deltah} \delta H_\xi = \Omega_{\Sigma}(g; \delta g,
\mathcal{L}_\xi g),
\end{equation}
for an arbitrary $\delta g$ which is tangent to field space
$\mathcal{F}$ at point $g$ of the solution subspace
$\overline{\mathcal{F}}$. One can put it in the
form~\cite{Iwald95}
\begin{equation}
\label{deltahnoether} \delta H_\xi = \int_\Sigma \xi^a \delta
\mathbf{C}_a + \int_{\partial \Sigma} [\delta \mathbf{Q} - \xi
\cdot \mathbf{\Theta}],
\end{equation}
where
\begin{equation}
\label{noethercharge} \mathbf{Q}_{a_1 \dots a_{n-2}} =
-\frac{1}{16\pi G}(\nabla^b \xi^c) \mbox{{\boldmath
$\epsilon$}}_{bca_1 \dots a_{n-2}}.
\end{equation}
If the equations of motion hold, then $\mathbf{C}_a = 0$, i.e.,
$\mathbf{C}_a$ correspond to ``constraints'' of the theory.
Equation (\ref{noethercharge}) defines the {\it Noether charge
$(n-2)$-form}, $ \mathbf{Q}$.  If $\delta g$ is tangent to
$\overline{\mathcal{F}}$, or satisfies linearized field equations,
then (\ref{deltahnoether}) becomes
\begin{equation}
\delta H_\xi = \int_{\partial \Sigma} [\delta \mathbf{Q} - \xi
\cdot \mathbf{\Theta}].
\end{equation}
It was shown in~\cite{Zwald} that if $\mbox{{\boldmath
$\omega$}}=0$ on $\mathcal{B}$ (case I), or $\xi$ is everywhere
tangent to the cross section $\partial \Sigma$ in $\mathcal{B}$
(case II), then ``conserved charges" $H_{\xi}$ exist, and in case
I, they are really conserved.

It should be noted, in one connected component of
$\overline{\mathcal{F}}$, that the conserved charge (\ref{deltah})
is uniquely defined up to a constant. Usually, we can choose a
natural ``reference solution" $g_0 \in \overline{\mathcal{F}}$ so
that this constant, $H_\xi[g_0]$, vanishes. Integrating the
variation parameter from $0$ to $\bar{\lambda}$ (which corresponds
to the solution we want to define the conserved charges.), the
conserved charge $H_\xi[g_{\bar{\lambda}}]$ is given by
\begin{equation}
H_\xi [g_{\bar{\lambda}}] = \int_0^{\bar{\lambda}} d \lambda
\int_{\partial \Sigma} [\delta \mathbf{Q}_{\lambda} - \xi \cdot
\mathbf{\Theta}_{\lambda}].
\end{equation}
Just as pointed out by the authors of the paper~\cite{Zwald}, this
definition does not depend on the choice of the paths connecting
$g_0$ and $g_{\bar{\lambda}}$. So the conserved charges are well
defined. In the next section, we will use the definition
 and give the conserved charges in AlAdS space-times.

In the remainder of this section, we will give some preliminary
analysis about the neighborhood of boundary of the AlAdS
space-times. The $n$-dimensional AlAdS space-times are solutions
of Einstein's equations with a negative cosmological constant,
whose Riemann tensor asymptotically approaches to that of exact
AdS space-time. A simple class of AlAdS space-times is AAdS
space-times which have boundary topology $\mathbf{R} \times
\mathbf{S}^{n-2}$. For general AlAdS space-times their boundary
topology may be different from the topology $\mathbf{R} \times
\mathbf{S}^{n-2}$.

\vspace{.1in} Let $(M,g_{\lambda}), \lambda \in \mathbf{R}$, be a
smooth one-parameter family of $n$-dimensional AlAdS space-times
which pass through the point $g_{\bar{\lambda}}$ (which
corresponds to the AlAdS space-time under discussion) in the
solution subspace $\overline{\Gamma}$. We assume that they have
the same conformal infinity. That is to say \noindent (i) one can
attach a boundary $\mathcal{B}$ to $M$ such that $\widetilde{M} =
M \cup \mathcal{B}$ is a manifold with boundary. For example $
\mathcal{B}\cong \mathbf{R} \times \mathbf{\Im} ^{n-2}$ and
$\mathcal{\mathbf{\Im}}^{n-2}$ denotes an $(n-2)$-dimensional
manifold whose topology may not be that of a round sphere
$\mathbf{S}^{n-2}$; \noindent (ii) on $\widetilde{M}$, there is a
family of smooth metrics $\bar{g}_{\lambda}$ and a smooth function
$\Omega$ (does not depend $\lambda$) such that $g_{\lambda} =
\Omega^{-2} \bar{g}_{\lambda}$, and such that $\Omega = 0$,
\begin{equation}
\label{domega} d \Omega \neq 0,
\end{equation}
at points of $\mathcal{B}$. The metrics on $\mathcal{B}$ induced
by $\bar{g}_{\lambda}$ are of the same form for all $\lambda$, and
can be denoted as $h$. For example, for the Schwarzschild-AdS
black holes with mass parameter as a variation parameter, their
metrics can be expressed as
\begin{eqnarray}
ds^2_{\lambda}&=&-(k-\frac{2\lambda}{r^{n-3}}+\frac{r^2}{\ell^2})dt^2
+\frac{dr^2}{k-\frac{2\lambda}{r^{n-3}}+\frac{r^2}{\ell^2}}+r^2d\sigma_{n-2}^2,
\end{eqnarray}
where $d\sigma^2_{n-2}$ denotes for the metric of
$(n-2)$-dimensional Einstein space with constant curvature
$(n-2)(n-3)k$ . In the case of $k=1$, one can choose a conformal
factor and boundary as $\Omega=\frac{\ell}{r}$, $\mathcal{B}={\bf
R} \times {\bf S}^{n-2}$,  and realize the completion described
above. The boundary metric $h$ is an Einstein static unverse with
radius $\ell$
\begin{equation}
d\bar{s}^2_{\lambda}|_{\mathcal{B}}  =
\Omega^2ds^2_{\lambda}|_{\mathcal{B}}   = -dt^2+\ell^2
d\sigma_{n-2}^2.
\end{equation}
For $h$ and each $\lambda$, in the neighborhood of $\mathcal{B}$,
there exists a unique conformal factor $\rho_{\lambda}$ or
coordinates $x_{\lambda}=(\rho_{\lambda},y)$ in which the metric
takes the form~\cite{Graham, Anderson}
\begin{eqnarray}
\label{coord}
\rho_{\lambda}^2 g_{\lambda}&=&\widetilde{g}_{\lambda} = d\rho_{\lambda}^2
+ \widetilde{h}_{\rho_{\lambda}} \nonumber \\
\widetilde{h}_{\rho_{\lambda}} &=& \widetilde{h}_{0} +
\rho_{\lambda} (\widetilde{h}_{\lambda})_{1}+ \cdots
+\rho_{\lambda}^{n-1} (\widetilde{h}_{\lambda})_{n-1} +
(\alpha_{\lambda})_{n-1} \rho_{\lambda}^{n-1}
\ln\rho_{\lambda}^2 + \cdots
\end{eqnarray}
where $\widetilde{h}_{\rho_{\lambda}}$ is chosen such that
$\widetilde{h}_{\rho_{\lambda}=0}$ is equal to the metric
$\widetilde{h}_0=h$ on $\mathcal{B}$, and in the neighborhood of
the boundary ${\mathcal B}$,
\begin{equation}
 \label{noshift} (\widetilde{h}_{\rho_{\lambda}})_{ab}
 {}^{(\lambda)}\widetilde{\nabla}^b \rho_{\lambda}= 0, \quad
(\widetilde{g}_{\lambda})^{ab} {}^{(\lambda)}\widetilde{\nabla}_a
\rho_{\lambda} {}^{(\lambda)}\widetilde{\nabla}_b \rho_{\lambda} =
1,
\end{equation}
where ${}^{(\lambda)}\widetilde{\nabla}_a$ is the covariant
derivative associated to $\widetilde{g}_{\lambda}$.
$(\widetilde{h}_{\rho_{\lambda}})_{ab}$ is the induced metric on
the surfaces $\mathcal{B}_{\rho_{\lambda}}$, the time-like
surfaces of constant $\rho_{\lambda}$ with coordinate $y$. In
fact, $\rho_{\lambda}$ is just the distance from the point to the
boundary. Here, for simplicity, we have set the AdS radius $\ell
=1$.

Straightforward computation shows that the Riemann tensor of
(\ref{coord}) is of the form of exact AdS up to a correction of
order $\rho_{\lambda}^{-3}$~\cite{Graham, SdHaro}. The asymptotic
analysis  reveals  that all coefficients shown in (\ref{coord})
except the traceless and divergenceless part of
$(\widetilde{h}_{\lambda})_{n-1}$ are locally determined in terms
of boundary data. So, $(\widetilde{h}_{\lambda})_{j}$, for $j\leq
n-2$ are independent of $\lambda$.  The logarithmic term appears
only when $n$ is odd, if one considers the pure gravity case. This
term is important in the context of AdS/CFT in odd dimensional
AlAdS space-times, which reflects the anomaly in the
even-dimensional dual conformal field theories~\cite{HS}.

Assume that the coordinates of point $p\in \epsilon \times
\mathcal{B}$ (The neighborhood of $\mathcal{B}$ has a direct
product form, we denote it by  $\epsilon \times \mathcal{B}$,
where $\epsilon$ is a small quantity.) are
$x_{\lambda}(p)=(\rho_{\lambda}(p),y(p))$. Consider
differomorphism $\phi_{\sigma}$ of $\widetilde{M}$ which has a
restriction on the neighborhood of $\mathcal{B}$
\begin{equation}
\phi_{\sigma}: \ \epsilon \times  \mathcal{B} \ \rightarrow \ \
\epsilon \times \mathcal{B}, \quad p \ \mapsto \phi_{\sigma}(p)
\end{equation}
with $\phi_{\sigma}(p)$ satisfying
\begin{equation}
x_{\lambda}(\phi_{\sigma}(p))=
(\rho_{\lambda+\sigma}(p),y(p)), \ \quad \forall \ \lambda \in
\mathbf{R}
\end{equation}
where $\epsilon$ is small enough such that these coordinates are
well defined in $\epsilon \times \mathcal{B} $ for all $\lambda$.
Then $\phi_{\sigma}$  forms a one parameter transformation on
$\epsilon \times \mathcal{B}$. If the vector field which generates
$\phi_{\sigma}$ is denoted by $\zeta$, then we have
\begin{eqnarray}
\mathcal{L}_{\zeta}\widetilde{g}_{\lambda}(p)
&=&\lim_{\sigma\rightarrow
0}\frac{1}{\sigma}\left[(\phi_{\sigma}^{\ast}
\widetilde{g}_{\lambda})-\widetilde{g}_{\lambda}
\right](p) \nonumber \\
&=&\lim_{\sigma \rightarrow 0}
\frac{1}{\sigma}\left[\phi_{\sigma}^{\ast}(d\rho_{\lambda}^2
+\widetilde{h}_{\rho_{\lambda}})
-(d\rho_{\lambda}^2+\widetilde{h}_{\rho_{\lambda}})\right](p)\nonumber \\
&=& \lim_{\sigma \rightarrow 0}
\frac{1}{\sigma}\left[\left(d(\rho_{\lambda}\circ\phi_{\sigma})^2
+\phi_{\sigma}^{\ast}\widetilde{h}_{\rho_{\lambda}}\right)
-\left(d\rho_{\lambda}^2+\widetilde{h}_{\rho_{\lambda}}\right)
\right](p)\nonumber \\
&=&\lim_{\sigma \rightarrow 0}
\frac{1}{\sigma}\left[\left(d\rho_{\lambda+\sigma}^2
+\widetilde{h}_{\rho_{\lambda+\sigma}}\right)
-\left(d\rho_{\lambda}^2+\widetilde{h}_{\rho_{\lambda}}\right)
\right](p)\nonumber \\
&=& \left[2d \left(\frac{d\rho_{\lambda}}{d\lambda}\right)d\rho_{\lambda}
+\frac{\partial \widetilde{h}_{\rho_{\lambda}}}{\partial
\rho_{\lambda}}\frac{d\rho_{\lambda}}{d\lambda}
\right](p)\nonumber
\end{eqnarray}
Thus, we have
\begin{equation}
\label{lied} \mathcal{L}_{\zeta}\widetilde{g}_{\lambda}=2d
\left(\frac{d\rho_{\lambda}} {d\lambda}\right)d\rho_{\lambda}
+\frac{\partial \widetilde{h}_{\rho_{\lambda}}}{\partial
\rho_{\lambda}}\frac{d\rho_{\lambda}}{d\lambda}.
\end{equation}
It should be noted that the restriction of $\phi_{\sigma}$ on
$\mathcal{B}$ is an identity. So, the vector field $\zeta$ must
vanish on $\mathcal{B}$ if it generates $\phi_{\sigma}$ in
$\epsilon \times \mathcal{B}$, i.e., it corresponds to a gauge
freedom according to the asymptotic symmetry transformations.
Consider the variation of $\widetilde{g}_{\lambda}$ at
$\bar{\lambda}$
\begin{equation}
\frac{d}{d\lambda}\widetilde{g}_{\lambda}|_{\bar{\lambda}}=
\frac{d}{d\lambda}(d\rho_{\lambda}^2+\widetilde{h}_{\rho_{\lambda}})|_{\bar{\lambda}}.
\end{equation}
With the help of (\ref{coord}) and (\ref{lied}), in the even
dimensional case, one has
\begin{eqnarray}
\label{varymetric} \frac{d}{d\lambda} \widetilde{g}_{\lambda}|_{\bar{\lambda}}&=&\left[
2d \left(\frac{d\rho_{\lambda}}{d\lambda}\right)d\rho_{\lambda}+\frac{\partial
\widetilde{h}_{\rho_{\lambda}}}{\partial
\rho_{\lambda}}\frac{d\rho_{\lambda}}{d\lambda}\right]_{\bar{\lambda}}
+\rho_{\lambda}^{n-1}\frac{d}{d\lambda}(\widetilde{h}_{\lambda})_{n-1}|_{\bar{\lambda}}
+O(\rho_{\bar{\lambda}}^{n-1})\nonumber \\
&=&\mathcal{L}_{\zeta}\widetilde{g}_{\bar{\lambda}}
+\rho_{\bar{\lambda}}^{n-1}\frac{d}{d\lambda}(\widetilde{h}_{\lambda})_{n-1}|_{\bar{\lambda}}
+O(\rho_{\bar{\lambda}}^{n-1}).
\end{eqnarray}
Thus, if the change of $\rho_{\lambda}$ with $\lambda$ gives a
contribution to the variation of $g_{\bar{\lambda}}$, then this
contribution is just a Lie-derivative of $g_{\bar{\lambda}}$ about
one vector field which vanishes on the boundary $\mathcal{B}$.
This equation is very useful in  our analysis below. The same
derivation will be done for giving the variation of the electric
part of Weyl tensor in the next section.

\section{Conserved Charges in Even Dimensional AlAdS Space-times}

To give the explicit form of conserved charges in AlAdS
space-times, we need to analyze the field equations. Our analysis
here is similar to that of Hollands {\it et al.}~\cite{Hollands}.
The difference is that we are treating even dimensional AlAdS
space-times instead of the AAdS space-times considered
in~\cite{Hollands}. Therefore the result is modified so that we
can calculate the conserved charges of more general solutions to
which the method of Ashtekar {\it et al}~\cite{AD} is not
applicable. The reader who wants to know more details of this
procedure may refer to the paper~\cite{Hollands}.

\subsection{Analysis of Einstein Equations}
To analyze Einstein's equations, we follow \cite{Hollands} and
introduce the tensor field for each $\lambda$ in the AlAdS
solution family $g_{\lambda , ab}$ described in the previous
section
\begin{equation}
(\widetilde{S}_{\lambda})_{ab} = \frac{2}{n-2}
(\widetilde{R}_{\lambda})_{ab} - \frac{1}{(n-1)(n-2)}
\widetilde{R}_{\lambda} (\widetilde{g}_{\lambda})_{ab},
\label{Sdef}
\end{equation}
In terms of this field and the coordinates of (\ref{coord}), in
the neighborhood of $\mathcal{B}$, Einstein's equations can be
rewritten as
\begin{equation} (\widetilde{S}_{\lambda})_{ab} = - 2
\rho_{\lambda}^{-1}  \ {}^{(\lambda)}\widetilde{\nabla}_a
(\widetilde{n}_{\lambda})_{b}.
\end{equation}
Treating $\rho_{\lambda}$ as ``time", one can obtain the equations
of ``constraint'' and ``evolution'' by using standard ADM
decomposition. The constraint equations are
\begin{eqnarray}
\label{restriction} - \widetilde{\mathcal{R}}_{\lambda} -
(\widetilde{K}_{\lambda})_{ab}(\widetilde{K}_{\lambda}){}^{ab} +
\widetilde{K}_{\lambda}^2 + 2(n-2)
\rho_{\lambda}^{-1}\widetilde{K}_{\lambda}
&=& 0 \,, \\
^{(\lambda)}\widetilde{D}^a (\widetilde{K}_{\lambda})_{ab} - \
^{(\lambda)}\widetilde{D}_b \widetilde{K}_{\lambda} &=& 0 \,,
\end{eqnarray}
where $^{(\lambda)}\widetilde{D}_a$ is the derivative operator
associated with $(\widetilde{h}_{\lambda})_{ab}$,
 $(\widetilde{K}_{\lambda})_{ab} = - (\widetilde{h}_{\lambda})_{a}
 {}^c (\widetilde{h}_{\lambda})_{b}{}^d  \ ^{(\lambda)}\widetilde{\nabla}_c
 (\widetilde{n}_{\lambda})_{d}$
is the extrinsic curvature of the surfaces
$\mathcal{B}_{\rho_{\lambda}}$ (with respect to the unphysical
metric). Here we have denoted
$(\widetilde{h}_{\rho_{\lambda}})_{ab}$ by
$(\widetilde{h}_{\lambda})_{ab}$ for simplicity, and
$\widetilde{\mathcal{R}}_{\lambda}$ is the intrinsic Ricci scalar
of $\mathcal{B}_{\rho_{\lambda}}$. The evolution equations are
\begin{eqnarray}
 \label{evol1} \frac{d}{d\rho_{\lambda}}( \widetilde{K}_{\lambda})_{a}{}^b
 &=& (\widetilde{\mathcal{R}}_{\lambda})_{a}{}^b
  + \widetilde{K}_{\lambda} (\widetilde{K}_{\lambda})_{a}{}^b +
\rho_{\lambda}^{-1}(n-2) (\widetilde{K}_{\lambda})_{a}{}^b
+ \rho_{\lambda}^{-1} \widetilde{K}_{\lambda} \delta_a{}^b, \\
\label{evol2} \frac{d}{d\rho_{\lambda}}
(\widetilde{h}_{\lambda})_{ab} &=& -2
(\widetilde{h}_{\lambda})_{bc} (\widetilde{K}_{\lambda})_{a}{}^c.
\end{eqnarray}
By assumption, $\mathcal{B}$ is a smooth boundary, which implies
that the fields $(\widetilde{h}_{\lambda})_{ab}$ and $
(\widetilde{K}_{\lambda})_{ab}$ must be smooth in a neighborhood
of $\mathcal{B}$. Consequently, multiplying the first evolution
equation by $\rho_{\lambda}$ and evaluating on $\mathcal{B}$, one
can immediately get
\begin{equation}
\label{boundaryzero} (\widetilde{K}_{\lambda})_{ab} |_ \mathcal{B}
= 0 = \frac{d}{d\rho_{\lambda}} (\widetilde{h}_{\lambda})_{ab} |_
\mathcal{B}.
\end{equation}

To investigate more systematically the consequences implied by
Eq.~(\ref{evol1}) and (\ref{evol2}), we express them in terms of
the traceless part $(\widetilde{p}_{\lambda})_{a}{}^b$ of
$(\widetilde{K}_{\lambda})_{a}{}^b$ and use the familiar
technique--Fefferman-Graham expansion~\cite{fg}:
\begin{eqnarray}
\label{fgexpansion} (\widetilde{h}_{\lambda})_{ab} &=&
\left(\widetilde{h}_{ab}\right)_{0} + \rho_{\lambda}
\left((\widetilde{h}_{\lambda})_{ab}\right)_{1}+ \cdots +
\rho_{\lambda}^{n-1}
\left((\widetilde{h}_{\lambda})_{ab}\right)_{n-1}
+\rho_{\lambda}^{n}\left((\widetilde{h}_{\lambda})_{ab}\right)_{n} +\cdots, \nonumber \\
(\widetilde{p}_{\lambda})_{a}{}^b&=&
\left((\widetilde{p}_{\lambda})_{a}{}^b \right)_{0}+
\rho_{\lambda} \left((\widetilde{p}_{\lambda})_{a}{}^b
\right)_{1}+ \cdots + \rho_{\lambda}^{n-1}
\left((\widetilde{p}_{\lambda})_{a}{}^b \right)_{n-1}
+\rho_{\lambda}^{n} \left((\widetilde{p}_{\lambda})_{a}{}^b
\right)_{n}+\cdots.
\end{eqnarray}
The logarithmic terms have not been included because we consider
even dimensional cases only, where each tensor
$\left((\widetilde{h}_{\lambda})_{ab}\right)_{j},
\,\left((\widetilde{p}_{\lambda})_{a}{}^b \right)_{j} $ are
independent of $\rho_{\lambda}$ in the sense that the
Lie-derivative along $(\widetilde{n}_{\lambda})^a$ vanishes.
Substituting the above expansion into~(\ref{evol1}) and
(\ref{evol2}), one can obtain the following recursion relations
\begin{eqnarray}
(n-2-j)\left((\widetilde{p}_{\lambda})_{a}{}^b \right)_{j} &=&
\left((\widetilde{\mathcal{R}}_{\lambda})_{a}{}^b\right)_{j-1} -
\frac{1}{n-1}\left(\widetilde{\mathcal{R}}_{\lambda}\right)_{j-1}
\delta_a{}^b \nonumber
\\  && {}\,
 - \sum_{m=0}^{j-1} \left(\widetilde{K}_{\lambda}\right)_{m}
 \left((\widetilde{p}_{\lambda})_{a}{}^b \right)_{j-1-m} \,,
\label{recursionp}
\\
(2n-3-j)\left(\widetilde{K}_{\lambda}\right)_{j} & =&
\left(\widetilde{\mathcal{R}}_{\lambda}\right)_{j-1} -
\sum_{m=0}^{j-1} \left(\widetilde{K}_{\lambda}\right)_{m}
\left(\widetilde{K}_{\lambda}\right)_{j-1-m} \,,
\label{recursionk}
\end{eqnarray}
and
\begin{equation} j \left((\widetilde{h}_{\lambda})_{ab}\right)_{j} =
-2\sum_{m=0}^{j-1}\left[
\left((\widetilde{h}_{\lambda})_{bc}\right)_{m}
\left((\widetilde{p}_{\lambda})_{a}{}^b \right)_{j-1-m}+
\frac{1}{n-1} \left((\widetilde{h}_{\lambda})_{ab}\right)_{m}
\left(\widetilde{K}_{\lambda}\right)_{j-1-m} \right] .
\label{recursionh}
\end{equation}
The ``initial conditions'' are, from Eq.~(\ref{boundaryzero}),
\begin{equation}
\left((\widetilde{p}_{\lambda})_{a}{}^b \right)_{0} =
\left(\widetilde{K}_{\lambda}\right)_{0} = 0,
\end{equation}
and $(\widetilde{h}_{ab})_{0}= h_{ab}$ is the metric of the
boundary $\mathcal{B}$. The key point of these equations is that
$\left((\widetilde{h}_{\lambda})_{ab}\right)_{j}$ and
$\left((\widetilde{K}_{\lambda})_{a}{}^b \right)_{l}$ are uniquely
determined in terms of the initial conditions for $j<n-1$ and
$l<n-2$. Therefore they are independent of $\lambda$. Thus, we
have
\begin{equation}
\label{varyh}
\frac{d}{d\lambda}(\widetilde{h}_{\lambda})_{ab}|_{\bar{\lambda}}=
\left (\frac{\partial (\widetilde{h}_{\lambda})_{ab}}{\partial
\rho_{\lambda}}\frac{d\rho_{\lambda}}{d\lambda}\right
)_{\bar{\lambda}}
+\rho_{\bar{\lambda}}^{n-1}\frac{d}{d\lambda}\left((\widetilde{h}_{\lambda})_{ab}\right)_{n-1}|_{\bar{\lambda}}
+O(\rho_{\bar{\lambda}}^n).
\end{equation}
As a result, any quantity that depends only on
$\left((\widetilde{h}_{\lambda})_{ab}\right)_{j}$ and
$\left((\widetilde{K}_{\lambda})_{a}{}^b \right)_{l}$ in range
$j<n-1$ and $l<n-2$, must be automatically independent of
$\lambda$.

The analysis of the recursion relation~(\ref{recursionp}) for
$j=n-2$ and constraint equation tell us that~\cite{Hollands}, once
the traceless, symmetric tensor
$\left((\widetilde{p}_{\lambda})_{a}{}^b \right)_{n-2}$ with the
divergence (determined by the constraint equations) is given, all
tensors $\left((\widetilde{p}_{\lambda})_{a}{}^b \right)_{j}$ and
$\left((\widetilde{h}_{\lambda})_{ab}\right)_{j}$ are uniquely
determined for $j \geq n-1$ via the evolution and constraint
equations. Thus, this tensor carries the full information about
the metric $(\widetilde{g}_{\lambda})_{ab}$ which is not already
supplied by the boundary conditions, i.e., the ``non-kinematical''
information. The tensor $\left((\widetilde{p}_{\lambda})_{a}{}^b
\right)_{n-2}$ is related to the electric part of the unphysical
Weyl tensor, as we will show  shortly. From the definition of the
tensor field $(\widetilde{S}_{\lambda})_{ab}$, we have
\begin{equation}
\label{RCS} (\widetilde{R}_{\lambda})_{abcd} =
(\widetilde{C}_{\lambda})_{abcd} + (\widetilde{g}_{\lambda})_{a[c}
(\widetilde{S}_{\lambda})_{d]b} - (\widetilde{g}_{\lambda})_{b[c}
(\widetilde{S}_{\lambda})_{d]a} \,.
\end{equation}
Using Einstein's equations, the definition of extrinsic curvature
and the Gauss-Coddazi relation, we have
\begin{equation}
(\widetilde{C}_{\lambda})^a{}_{bcd} (\widetilde{n}_{\lambda})^b
(\widetilde{n}_{\lambda})^d= - (\widetilde{K}_{\lambda})^a{}_{b}
(\widetilde{K}_{\lambda})^b{}_{c} +
{\mathcal{L}_{\widetilde{n}_{\lambda}}}
(\widetilde{K}_{\lambda})^a{}_{c}
  - \rho_{\lambda}^{-1} (\widetilde{K}_{\lambda})^a{}_{c} \, .
\label{A-Key-equation-2} \end{equation} This equation can be
expanded in powers of $\rho_{\lambda}$ like the metric. We thereby
obtain equations for the expansion coefficients. At the order
$n-3$, one has the relation
\begin{equation}
\label{weylrecurs} \frac{1}{n-3}
\left((\widetilde{C}_{\lambda})_{acbd} (\widetilde{n}_{\lambda})^c
(\widetilde{n}_{\lambda})^d\right)_{n-3} =
\left((\widetilde{K}_{\lambda})_{ab}\right)_{n-2} - \frac{1}{n-3}
\sum_{m=0}^{n-3} \left((\widetilde{K}_{\lambda})_{ac}\right)_m
\left((\widetilde{K}_{\lambda})_{b}{}^c\right)_{n-3-m} \,.
\end{equation}
However, the coefficients appearing in the sum are independent of
$\lambda$, and they are determined by boundary data. Consequently,
we can obtain the variation relation
\begin{equation}
\label{varycn-3} \frac{d}{d\lambda}\left[\frac{1}{n-3}
\left((\widetilde{C}_{\lambda})_{acbd} (\widetilde{n}_{\lambda})^c
(\widetilde{n}_{\lambda})^d\right)_{n-3}\right]_{\bar{\lambda}}
=\frac{d}{d\lambda}\left[\left((\widetilde{K}_{\lambda})_{ab}\right)_{n-2}\right]_{\bar{\lambda}}.
\end{equation}
It is easy to see, at order $j<n-3$, the coefficients of
$(\widetilde{C}_{\lambda})_{abcd} (\widetilde{n}_{\lambda})^b
(\widetilde{n}_{\lambda})^d$ are independent of $\lambda$ because
they are fixed by the boundary data as we mentioned above.
 Thus, the variation of these coefficients are zero, i.e.,
\begin{equation}
\label{cjsmallthannmthree} \frac{d}{d\lambda}\left[
\left((\widetilde{C}_{\lambda})_{acbd} (\widetilde{n}_{\lambda})^c
(\widetilde{n}_{\lambda})^d\right)_{j}\right]_{\bar{\lambda}}=0.
\end{equation}
Combining (\ref{recursionh}) with (\ref{varycn-3}), we get
\begin{equation}
\frac{d}{d\lambda}\left[\frac{1}{n-3}
\left((\widetilde{C}_{\lambda})_{acbd} (\widetilde{n}_{\lambda})^c
(\widetilde{n}_{\lambda})^d\right)_{n-3}\right]_{\bar{\lambda}}
=\frac{d}{d\lambda}\left[-\frac{n-1}{2}\left((\widetilde{h}_{\lambda})_{ab}\right)_{n-1}\right]_{\bar{\lambda}}.
\end{equation}
Substituting this result into (\ref{varyh}), we immediately have the following equation
\begin{equation}
\label{varyhc}
\frac{d}{d\lambda}(\widetilde{h}_{\lambda})_{ab}|_{\bar{\lambda}}=\left[\frac{\partial
(\widetilde{h}_{\lambda})_{ab}}{\partial
\rho_{\lambda}}\frac{d\rho_{\lambda}}{d\lambda}\right]_{\bar{\lambda}}
-\frac{2}{n-1}\rho_{\bar{\lambda}}^{n-1}\frac{d}{d\lambda}\left[\frac{1}{n-3}
\left((\widetilde{C}_{\lambda})_{acbd} (\widetilde{n}_{\lambda})^c
(\widetilde{n}_{\lambda})^d\right)_{n-3}\right]_{\bar{\lambda}}+O(\rho_{\bar{\lambda}}^n).
\end{equation}
The similar procedure to deduce Eq. (\ref{varymetric}) can be used
to $(\widetilde{C}_{\lambda})_{acbd} (\widetilde{n}_{\lambda})^c
(\widetilde{n}_{\lambda})^d$, and once again one gets that the
variation of $(\widetilde{C}_{\lambda})_{acbd}
(\widetilde{n}_{\lambda})^c (\widetilde{n}_{\lambda})^d$ can be
divided into two parts (In fact, any quantity which can be
expanded as (\ref{fgexpansion}) always has such a variation
relation)
\begin{eqnarray}
\frac{d}{d\lambda}\left[(\widetilde{C}_{\lambda})_{acbd}
(\widetilde{n}_{\lambda})^c
(\widetilde{n}_{\lambda})^d\right]_{\bar{\lambda}}
&=&\mathcal{L}_{\zeta}\left[(\widetilde{C}_{\bar{\lambda}})_{abcd}
(\widetilde{n}_{\bar{\lambda}})^b
(\widetilde{n}_{\bar{\lambda}})^d\right] \nonumber
\\
&&+\rho_{\bar{\lambda}}^{n-3}\frac{d}{d\lambda}\left[\left((\widetilde{C}_{\lambda})_{abcd}
(\widetilde{n}_{\lambda})^b
(\widetilde{n}_{\lambda})^d\right)_{n-3}\right]_{\bar{\lambda}}+O(\rho_{\bar{\lambda}}^{n-2}).
\end{eqnarray}
Multiplying this equation by $\rho_{\bar{\lambda}}^{3-n}$ and
considering
$\mathcal{L}_{\zeta}\rho_{\bar{\lambda}}=\frac{d\rho_{\lambda}}{d\lambda}|_{\bar{\lambda}}$,
we arrive at
\begin{eqnarray}
\label{varyeleweyl}
\frac{d}{d\lambda}\left[\rho_{\lambda}^{3-n}(\widetilde{C}_{\lambda})_{acbd}
(\widetilde{n}_{\lambda})^c
(\widetilde{n}_{\lambda})^d\right]_{\bar{\lambda}}
&=&\mathcal{L}_{\zeta}\left[\rho_{\bar{\lambda}}^{3-n}(\widetilde{C}_{\bar{\lambda}})_{acbd}
(\widetilde{n}_{\bar{\lambda}})^c
(\widetilde{n}_{\bar{\lambda}})^d\right]\nonumber \\
&&+\frac{d}{d\lambda}\left[\left((\widetilde{C}_{\lambda})_{acbd}
(\widetilde{n}_{\lambda})^c
(\widetilde{n}_{\lambda})^d\right)_{n-3}\right]_{\bar{\lambda}}+O(\rho_{\bar{\lambda}}).
\end{eqnarray}
Define the electric part of the unphysical Weyl tensor as
\begin{equation}
(\widetilde{E}_{\lambda})_{ab}=\frac{1}{n-3}
\rho_{\lambda}^{3-n}\left((\widetilde{C}_{\lambda})_{acbd}
(\widetilde{n}_{\lambda})^c (\widetilde{n}_{\lambda})^d\right).
\end{equation}
It should be noted here, although $(\widetilde{E}_{\lambda})_{ab}$
may be divergent when $\rho_{\lambda}$ approaches to zero, the
variation $\frac{d}{d\lambda}(\widetilde{E}_{\lambda})_{ab}$ is
always finite if one fixes the conformal factor as a gauge
condition, which can be understood from the above discussion. In
what follows, we will treat only the difference or variation of
this tensor under the gauge as we mentioned above. In that case,
the divergence will not appear. By using  Eq.(\ref{varyhc}) and
Eq.(\ref{varyeleweyl}), we have
\begin{eqnarray}
\frac{d}{d\lambda}(\widetilde{h}_{\lambda})_{ab}|_{\bar{\lambda}}&=&\left[\frac{\partial
(\widetilde{h}_{\lambda})_{ab}}{\partial
\rho_{\lambda}}\frac{d\rho_{\lambda}}{d\lambda}\right]_{\bar{\lambda}}
+\frac{2}{n-1}\rho_{\bar{\lambda}}^{n-1}\mathcal{L}_{\zeta}(\widetilde{E}_{\bar{\lambda}})_{ab}
-\frac{2}{n-1}\rho_{\bar{\lambda}}^{n-1}\frac{d}{d\lambda}(\widetilde{E}_{\lambda})_{ab}|_{\lambda}
+O(\rho_{\bar{\lambda}}^n).
\end{eqnarray}
At this stage, we deduce an important result in this section
\begin{equation}
\label{varyh1}
\frac{d}{d\lambda}(\widetilde{g}_{\lambda})_{ab}|_{\bar{\lambda}}
=\mathcal{L}_{\zeta}(\widetilde{g}_{\bar{\lambda}})_{ab}
+\frac{2}{n-1}\rho_{\bar{\lambda}}^{n-1}\mathcal{L}_{\zeta}(\widetilde{E}_{\bar{\lambda}})_{ab}
-\frac{2}{n-1}\rho_{\bar{\lambda}}^{n-1}\frac{d}{d\lambda}(\widetilde{E}_{\lambda})_{ab}|_{\bar{\lambda}}
+O(\rho_{\bar{\lambda}}^n).
\end{equation}
Thus the variation of $\widetilde{g}_{\lambda}$ can be splitted
into two parts: the first part is given by the dependence of
$\rho_{\lambda}$ on $\lambda$, and can be regarded as a result of
trivial differomorphism which is generated by the vector field
$\zeta$, and  the second part is given by the ``dynamical
information" of the system.  Therefore we will mainly concentrate
on the second part below.

\subsection{Derivation of Conserved Charges}

The parts related to the Lie-derivative in Eq.(\ref{varyh1})
correspond to the gauge freedoms, they can be gauge fixed away.
 We fix $\rho_{\lambda}$ to be $\rho=\rho_{\bar{\lambda}}$ such
that $\zeta=0$, and consider the variation of the form
\begin{equation}
\frac{d}{d\lambda}(\widetilde{g}_{\lambda})_{ab}|_{\bar{\lambda}}
=-\frac{2}{n-1}\rho^{n-1}\frac{d}{d\lambda}(\widetilde{E}_{\lambda})_{ab}|_{\bar{\lambda}}+O(\rho^n).
\end{equation}
Recalling  that $\rho$ has been regarded as a fixed function which
does not depend on $\lambda$, after integrating $\lambda$ from $0$
to $\bar{\lambda}$, we have
\begin{equation}
\label{gauge}
(\widetilde{g}_{\bar{\lambda}})_{ab}-(\widetilde{g}_{0})_{ab}=
-\frac{2}{n-1}\rho^{n-1}\left[(\widetilde{E}_{\bar{\lambda}})_{ab}
-(\widetilde{E}_{0})_{ab}\right]+O(\rho^n),
\end{equation}
where $\widetilde{g}_{0}$ has the same conformal infinity as
$\widetilde{g}_{\bar{\lambda}}$ and plays the role of the metric
of the reference background solution. Here we have already assumed
that $\widetilde{g}_{0}$ and $\widetilde{g}_{\bar{\lambda}}$
belong to the same connected component of the solution subspace;
therefore, there is a smooth path that connects them. We emphasize
that, although $(\widetilde{E}_{\lambda})_{ab}$ may be divergent
when $\rho $ approaches to zero, the difference
$\left[(\widetilde{E}_{\bar{\lambda}})_{ab}
-(\widetilde{E}_{0})_{ab}\right]$ is always finite once one fixes
the conformal factor to be $\rho$. In other words, the leading
order of the variation of $(\widetilde{E}_{\lambda})_{ab}$ and
$\left[(\widetilde{E}_{\bar{\lambda}})_{ab}
-(\widetilde{E}_{0})_{ab}\right]$ has the form $\rho^{3-n}$.

Following \cite{Hollands}, we view $\widetilde{g}_{\bar{\lambda}}$
with the expression (\ref{gauge}) as a ``gauge condition" on the
metric, i.e. as a particular representative in the equivalence
class of metrics which is diffeomorphic to $g_{\bar{\lambda}}$.
The (on-shell) metric variations respecting this gauge choice
(with $\rho_{\lambda}$ be fixed as $\rho$) therefore take the form
\begin{equation}
\frac{d}{d\lambda}
(g_{\lambda})_{ab}|_{\bar{\lambda}}=(\gamma_{\bar{\lambda}})_{ab}+\mathcal{L}_{\eta}(g_{\bar{\lambda}})_{ab},
\end{equation}
where the first piece $(\gamma_{\bar{\lambda}})_{ab}$ is a metric
variation of the form
\begin{equation}
(\gamma_{\bar{\lambda}})_{ab}=-\frac{2}{n-1}\rho^{n-1}\frac{d}{d\lambda}(\widetilde{E}_{\lambda})_{ab}|_{\bar{\lambda}}
+O(\rho^n),
\end{equation}
and the second piece is an infinitesimal diffeomorphism generated by
an arbitrary vector field $\eta$ respecting the gauge choice, i.e., a diffeo satisfying
$\mathcal{L}_{\eta}g_{0}\sim O(\rho^n)$, where $g_{0}$ is the metric of reference space-time.
Thus,
\begin{equation}
\mathcal{L}_{\eta}(g_{\bar{\lambda}})_{ab}
=-\frac{2}{n-1}\rho^{n-1}\mathcal{L}_{\eta}\left[(\widetilde{E}_{\bar{\lambda}})_{ab}
-(\widetilde{E}_{0})_{ab}\right]+O(\rho^n).
\end{equation}
Inserting these expressions into the definition of the symplectic
current form $\mbox{{\boldmath $\omega$}}(g, \delta_1 g, \delta_2
g)$, we see that $\mbox{{\boldmath $\omega$}}|_{\mathcal{B}}=0$.
Hence, the conserved charges $H_{\xi}$ exist and are indeed
conserved.

The variation of Noether charge $(n-2)$-form is~\cite{Hollands}
\begin{equation}
\label{delqu} \frac{d}{d\lambda} (\mathbf{Q}_{\lambda})_{a_1 \dots
a_{n-2}}|_{\bar{\lambda}} = \frac{1}{8 \pi G}
(\widetilde{\mbox{\boldmath $\epsilon$}}_{\bar{\lambda}})_{a_1
\dots a_{n-2}bc} (\widetilde{n}_{\bar{\lambda}})^b
\frac{d}{d\lambda} (\widetilde{E}_{\lambda})^c{}_{d}
\xi^d|_{\bar{\lambda}} + O(\rho) \,.
\end{equation}
Using the relation
\begin{equation}
^{(n)} {}\widetilde{\mbox{\boldmath $\epsilon$}}_{\lambda} =
 \widetilde{n}_{\lambda} \wedge {}^{(n-1)} \widetilde{\mbox{\boldmath $\epsilon$}} = \widetilde{n}_{\lambda} \wedge
\widetilde{u} \wedge {}^{(n-2)} \widetilde{\mbox{\boldmath
$\epsilon$}},
\end{equation}
among the $n$-dimensional volume form, the induced $(n-1)$-dimensional
volume form of the boundary $\mathcal{B}$ and the $(n-2)$-dimensional volume form of the cross section
$\partial\Sigma$, we can rewrite Eq.(\ref{delqu}) as
\begin{equation} \frac{d}{d\lambda} (\mathbf{Q}_{\lambda})_{a_1 \dots a_{n-2}} |_{\bar{\lambda}}= \frac{-1}{8 \pi G}
\frac{d}{d\lambda} \left[ {}^{(n-2)} \widetilde{\mbox{\boldmath
$\epsilon$}}_{a_1 \dots a_{n-2}} \,
(\widetilde{E}_{\lambda})^c{}_{d} \widetilde{u}_c \xi^d
\right]_{\bar{\lambda}} \,, \,\,\, \mbox{on $\mathcal{B}$}.
\end{equation}
A similar calculation can be done and gives
$\mathbf{\Theta}_{\bar{\lambda}} |_ \mathcal{B} = 0$. Thus, we
have
\begin{equation}
\frac{d}{d\lambda} H_{\xi}[g_{\lambda}]_{\bar{\lambda}} =
\int_{\partial\Sigma} \frac{d}{d\lambda}
\mathbf{Q}_{\lambda}|_{\bar{\lambda}}= \frac{-1}{8 \pi G} \,\,
\left[\frac{d}{d\lambda} \int_{\partial \Sigma}
(\widetilde{E}_{\lambda})_{ab} \widetilde{u}^b \xi^a \, d
\widetilde{S}\right]_{\bar{\lambda}}.
\end{equation}
Integrating this equation over $\lambda$ from $\lambda=0$ to
$\bar{\lambda}$, yields
\begin{equation}
H_{\xi}[g_{\bar{\lambda}}]-H_{\xi}[g_{0}]=\frac{-1}{8 \pi G} \,\,
\int_{\partial \Sigma} \left[(\widetilde{E}_{\bar{\lambda}})_{
ab}-(\widetilde{E}_{0})_{ ab}\right] \widetilde{u}^b \xi^a \, d
\widetilde{S}.
\end{equation}
Choosing $H_{\xi}[g_0]=0$ for all asymptotic symmetric
representatives $\xi^a$,  we get the result
\begin{eqnarray}
\label{conservedcharge} H_{\xi}[g_{\bar{\lambda}}]&=&\frac{-1}{8
\pi G} \,\,\int_{\partial
\Sigma}\left[(\widetilde{E}_{\bar{\lambda}})_{a
b}-(\widetilde{E}_{0})_{ab}\right] \widetilde{u}^b \xi^a \, d
\widetilde{S},
\end{eqnarray}
which can also be expressed by Weyl tensor
\begin{equation}
\label{cccon} H_{\xi}[g_{\bar{\lambda}}]=\frac{-1}{8 \pi G (n-3)}
\int_{\partial \Sigma} \lim _{\rightarrow
\mathcal{B}}\rho^{3-n}\left[(\widetilde{C}_{\bar{\lambda}})_{acbd}(\widetilde{n}_{\bar{\lambda}})^{c}
(\widetilde{n}_{\bar{\lambda}})^{d}-(\widetilde{C}_{0})_{acbd}(\widetilde{n}_{0})^{c}
(\widetilde{n}_{0})^{d}\right]\widetilde{u}^b \xi^a \, d
\widetilde{S}.
\end{equation}
This is nothing, but the one we obtained in this paper for the
expression of conserved charges for AlAdS space-times,
 This expression is conformal invariant, so in fact we can choose
a more simple conformal factor instead of $\rho$ (for example,
$\Omega=1/r$ in Sec.~2) to calculate the conserved charges. In
AAdS space-times, the reference space-time is fixed to be an exact
AdS space-time which is conformal flat, and in this case the above
formula (\ref{cccon}) reduces to the definition of Ashtekar {\it
et al}~\cite{AD}. For a general AlAdS space-time, however, one has
no prior background which can be chosen as  an appropriate
reference background  solution.

 Although we have discussed only the even
dimensional cases, it is easy to see that the above procedure can
be extended to the more general AlAdS cases where the
Fefferman-Graham expansion like (\ref{fgexpansion}) without log
terms can be implemented. For example, one can find that in the
static AlAdS space-time setting with Ricci flat boundary, for both
even and odd dimensional cases, no log terms arise in the
expansion~\cite{Galloway}. Thus, the same conserved charges for
those space-times can be defined as Eq.(\ref{cccon}).  As an
example, in Sec.~7 we will calculate conserved charges for
(un)warped black brane space-times, which belong to such a kind of
space-times.

\section{Taub-Nut-AdS and Taub-Bolt-AdS Space-times}

Taub-Nut-AdS and Taub-Bolt-AdS solutions are AlAdS solutions. For
example, in the Euclidean sector, their boundaries are $U(1)$
bundles over $\underbrace{S^2 \times \cdots \times S^2}_{k}$  for
$2k+2$ dimension\footnote{Where $S^2 \times \cdots \times S^2$ is
called base space. It may be other Einstein-K$\ddot{a}$hler
manifold, for example $T^2 \times \cdots \times T^2$, $S^2 \times
\cdots \times S^2 \times T^2 \times \cdots \times T^2$ and
$CP(k)$. We only consider the sphere cases in this paper. }, their
metrics can be expressed as~\cite{AAwad}
\begin{equation}
\label{tubnutgeneral} ds^{2}=F(r)(d\tau +2n\cos \theta _{i}d\phi
_{i})^2+\frac{dr^{2}}{F(r)} +(r^{2}-n^{2})\sum_{i=1}^{i=k}(d\theta
_{i}^{2}+\sin ^{2}\theta _{i}d\phi _{i}^{2}),
\end{equation}
where $i$ is summed from $1$ to $k$ and  $F(r)$ is given by
\begin{equation}
F(r)=\frac{r}{(r^{2}-n^{2})^{k}}\int^{r}\left[
\frac{(s^{2}-n^{2})^{k}}{ s^{2} }+\frac{2k+1}{\ell
^{2}}\frac{(s^{2}-n^{2})^{k+1}}{s^{2}}\right] ds- \frac{2Mr
}{(r^{2}-n^{2})^{k}}.
\end{equation}
One can find $F(r)\sim \frac{r^2}{\ell^2}$ when $r$ approaches to
infinity. If we choose the conformal factor as
$\Omega=\frac{\ell}{r}$, the boundary metrics have the form
\begin{equation}
d\overline{s}^2|_{\mathcal{B}}=\Omega^2 ds^2|_{\mathcal{B}} \sim
\left(d \tau + 2n \cos{\theta_i} d{\phi_i} \right)^2+ \ell^2
\left(d{\theta_i}^2+\sin^{2} \theta_i d {\phi_i}^2\right).
\end{equation}
It is easy to find that these boundaries are trivial bundles
(i.e., direct product manifolds $S^1\times S^2 \times \cdots
\times S^2$ ) when the Nut charges vanish. Therefore non-vanishing
Nut charges give rise to the non-triviality of the bundles.

Before calculating their conserved charges, the following points
are worthwhile to stress: \noindent (i) Note that the Nut charges
$n$ appear in the boundary metrics, we conclude that solutions
with different nut charges have different boundary metrics. Thus
we can not view a solution with Nut charge $n_1$ as a reference
solution for calculating conserved charges of another solution
with different Nut charge $n_2$. This is very different from the
AAdS case where the boundary is fixed to be Einstein static
space-time. \noindent (ii) Since our formula is background
dependent, there is some freedom to choose the reference solution.
 Hawking {\it et al.}~\cite{Hawking} and
Chamblin {\it et al.}~\cite{AChmblin} have argued that one can use
4-dimensional Taub-Nut-AdS solution as the reference solution,
when considering Taub-Bolt-AdS black hole solutions. They have
calculated the corresponding thermodynamic quantities of
4-dimensional Taub-Bolt-AdS solution by using ``background
subtraction" method. The resultant thermodynamic quantities obey
the first law of black hole thermodynamics. Using ``Noether
charge" method Clarkson {\it et al.} \cite{Mann1} have computed
the conserved charges ($Q_{\rm Noether}$) by treating the Nut
solutions as reference solutions in higher dimensional cases. They
also have computed these quantities ($M_{\rm bolt},M_{\rm Nut}$)
by using the ``boundary counterterm" method and found a relation
between them:
\begin{equation}
Q_{\rm Noether}=M_{\rm bolt}-M_{\rm Nut}
\end{equation}
Motivated by these works and to compare our method with other
methods, in this section, we will select the Nut solutions as
reference solutions. Then we calculate the conserved quantities
for $4-,6-,8-$ and $10$-dimensional cases by using our new formula
(\ref{cccon}). Our results agree with those given in \cite{Mann1}
in any dimension.  Furthermore, let us note that if chooses a
massless solution, but with the same Nut charge, as the reference
solution, one can also get finite results, as shown below. In that
case,  our results are the same as those resulting from the
``boundary counterterm" method. For simplicity, we will mainly
consider the Euclidean sector for these metrics. The quantities in
the Lorentzian sector can be obtained by analytically continuing
the coordinate $\tau$ and also the parameter $n$ (i.e., one
replaces $n^2$ with $-N^2$).

\subsection{Four Dimensional Solutions}
The four dimensional Nut charged AdS solution has the following
form~\cite{AAwad}
\begin{equation}
\label{tubnut4} ds^2 = F(r) (d \tau + 2n \cos{\theta} d{\phi} )^2+
F(r)^{-1} dr^2 + (r^2-n^2)(d{\theta}^2+\sin^{2} \theta d {\phi}^2),
\end{equation}
where $F(r)$ is given by
\begin{equation}
F(r)={1 \over \ell ^2(r^2-n^2)}\left[\ell ^2(r^2+n^2)-2Mr\ell
^2+(r^4-6n^2r^2-3n^4)\right].
\end{equation}
In order for this solution to describe a Nut solution, the mass
parameter has to be fixed to be
\begin{equation}
M_n=\frac{n(\ell^2-4n^2)}{\ell^2},
\end{equation}
so that $F(r = n) = 0$ and the dimension of the fixed-point set of
${\partial}_{\tau}$ is zero. Fixing the mass parameter at this
value, $F(r)$ is then given by
\begin{equation}
F_{n}(r)=\frac{r-n}{r+n}(1+\ell^{-2}(r-n)(r+3n)).
\end{equation}
On the other hand, the Bolt solution is given by
\begin{equation}
F(r)=F_{b}(r)=\frac{r^2-2M_{b}r+n^2+\ell^{-2}(r^4-6n^2r^2-3n^4)}{r^2-n^2},
\end{equation}
where
\begin{equation}
M_b =\frac{r_{b}^2+n^2}{2r_b}+\frac{1}{2\ell^2}(r_{b}^3-6n^2r_b
-3\frac{n^4}{r_b}),
\end{equation}
with
\begin{equation}
r_{b\pm}=\frac{\ell^2}{12n}(1\pm
\sqrt{1-48\frac{n^2}{\ell^2}+144\frac{n^4}{\ell^4}}).
\end{equation}
For $r_b$ to be real the discriminant must be no-negative.
Furthermore, the condition must be satisfied: $r_b > n$, which
 gives
\begin{equation}
n\leq (\frac{1}{6}-\frac{\sqrt{3}}{12})^{1/2} \ell.
\end{equation}
Treating the Nut solution as the reference solution, we  give the
 conserved charges of the Taub-Bolt-AdS solution, corresponding to the Killing vector
 $\partial_\tau$,
 as \footnote{Hereafter we use the orientation convention of Gibbons {\it et
al.} \cite{Gibbons} such that $ d \widetilde{S}_{t} $ is
positive.}
\begin{equation}
H_{\partial_{\tau}}[g_1]=\frac{\ell}{8 \pi G } \int_{\partial
\Sigma}
\Omega^{-1}\left[(\widetilde{C}_{1})_{acbd}(\widetilde{n}_1)^{c}
(\widetilde{n}_1)^{d}-(\widetilde{C}_{0})_{acbd}(\widetilde{n}_0)^{c}
(\widetilde{n}_0)^{d}\right]\widetilde{u}^b (\partial_{\tau})^a \,
d \widetilde{S},
\end{equation}
where  the subscripts ``$1$" and``$0$" correspond to ``Bolt"  and
``Nut" solution, respectively. By straightforward calculations, it
turns out that the leading order of the Weyl tensor for the
solution (\ref{tubnut4}) is
\begin{equation}
\widetilde{C}^{\tau}{}_{r\tau r}=C^{\tau}{}_{r\tau r}\sim \frac{2
M \ell^2}{r^5}+ O(\frac{1}{r^6}).
\end{equation}
Choosing the conformal factor as $\frac{\ell}{r}$, one then has
\begin{eqnarray}
(\widetilde{E}_{1})^{\tau}{}_{\tau}-(\widetilde{E}_0)^{\tau}{}_{\tau}&=&
\Omega^{-1}\left[(C_{1})^{\tau}{}_{ r\tau r}
(\widetilde{n}_1)^{r}(\widetilde{n}_1)^{r}-(C_{0})^{\tau}{}_{
r\tau r}(\widetilde{n}_0)^{r}(\widetilde{n}_0)^{r}\right]  \nonumber \\
&\sim&\frac{r^5}{\ell^5}\frac{2\ell^2 (M_{b}-M_{n})}{r^5}
\nonumber \\
&=&\frac{2 (M_{b}-M_{n})}{\ell^3}.
\end{eqnarray}
Note that $ d\widetilde{S}_{\tau}= \ell^2 \sin{\theta}d\theta $,
 we obtain the mass of the Taub-Bolt-AdS solution
\begin{eqnarray}
\Delta M &=&\frac{\ell}{8\pi G}\int_{\partial \Sigma}\frac{2
(M_{b}-M_{n})}{\ell^3}\ell^2 \sin{\theta}d\theta d\phi\\ \nonumber
&=&\frac{(M_{b}-M_{n})}{G}.
\end{eqnarray}
This is completely in agreement with  the one in \cite{AChmblin,
Mann1}. If one writes the metric in the form in~\cite{Hawking},
the same result as in~\cite{Hawking} can also be obtained.

\subsection{Six Dimensional Solutions}
The six dimensional Nut charged AdS solution has the
metric~\cite{AAwad}
\begin{eqnarray}
\label{tubnut6} ds^2 &=& F(r) (d \tau + 2n \cos{\theta_1}
d{\phi}_1 + 2n \cos{ \theta}_2d { \phi }_2)^2+ F(r)^{-1} dr^2
\nonumber \\
 &&+ (r^2-n^2)(d{\theta_1}^2+\sin^{2} \theta_{1} d
{\phi_1}^2+d{\theta_2}^2+\sin^{2} \theta_{2} d {\phi_2}^2),
\end{eqnarray}
where $F(r)$ is given by
\begin{equation}
F(r)={1 \over
3\ell^2(r^2-n^2)^2}\left[3r^6+(\ell^2-15n^2)r^4-3n^2(2\ell^2-15n^2)r^2-6Mr\ell^2-3n^4(\ell^2-5n
^2)\right].
\end{equation}
When the mass parameter $M$ is fixed to be
\begin{equation}
M_n={4n^3(6n^2-\ell^2) \over 3\ell^2},
\end{equation}
this solution describes a Nut solution with $F(r = n) = 0$, so
that the dimension of the fixed-point set of ${\partial}_{\tau}$
is zero. Fixing the mass at this value, $F(r)$ is changed to
\begin{equation}
F_{n}(r)={(r-n)(3r^3+9nr^2+(\ell^2+3n^2)r+3n(\ell^2-5n^2)) \over
3(r+n)^2\ell^2}.
\end{equation}
A regular Bolt solution has the mass parameter
\begin{equation}
M = M_b = {-1 \over
6\ell^2}[3{r_b}^5+(\ell^2-15n^2){r_b}^3-3n^2(2\ell^2-15n^2){r_b}-3n^4(\ell^2-5n^2)/r_b],
\end{equation}
where $r_b$ is a function of $n$ and $\ell$
\begin{equation}
r_{b \pm} = {1 \over 30 n}\left(\ell^2\pm
\sqrt{\ell^4-180n^2\ell^2+900n^4}\right).
\end{equation}
To have a real value of $r_b$ the discriminant in the above
equation must be non-negative. The condition $r_b > n$ leads to
\begin{equation}
n \le \left({3-2\sqrt{2} \over 30} \right)^{1\over 2}\ell.
\end{equation}
Treating the Nut solution as the reference solution, we can give
the conserved charges for the Bolt solution
\begin{equation}
H_{\partial_{\tau}}[g_1]=\frac{\ell}{8 \pi G \cdot 3}
\int_{\partial \Sigma}
\Omega^{-3}\left[(\widetilde{C}_{1})_{acbd}(\widetilde{n}_1)^{c}
(\widetilde{n}_1)^{d}-(\widetilde{C}_{0})_{acbd}(\widetilde{n}_0)^{c}
(\widetilde{n}_0)^{d}\right]\widetilde{u}^b (\partial_{\tau})^a \,
d \widetilde{S}.
\end{equation}
By straightforward calculation, the first two terms of the Weyl
tensor for the solution (\ref{tubnut6}) are
\begin{equation}
\label{eq82}
 \widetilde{C}^{\tau}{}_{r\tau r}=C^{\tau}{}_{r\tau
r}\sim \frac{4n^2(\ell^2-6n^2)}{r^6}+\frac{12 \ell^2 M}{r^7}+
O(\frac{1}{r^8}).
\end{equation}
Note that the first term on the right hand side of the above
equation (\ref{eq82}) is independent of the mass parameter $M$.
Thus if we directly apply the formula given by Ashtekar {\it et
al.}~\cite{AD} to this six dimensional Bolt solution, obviously we
will get a divergent result due to (\ref{eq82}). However, using
our formula (\ref{cccon}), we have
\begin{eqnarray}
(\widetilde{E}_{1})^{\tau}{}_{\tau}-(\widetilde{E}_0)^{\tau}{}_{\tau}&=&
\frac{1}{3}\Omega^{-3}\left[(C_{1})^{\tau}{}_{ r\tau r}
(\widetilde{n}_1)^{r}(\widetilde{n}_1)^{r}-(C_{0})^{\tau}{}_{
r\tau r}(\widetilde{n}_0)^{r}(\widetilde{n}_0)^{r}\right]  \nonumber\\
&\sim&\frac{r^7}{\ell^7}\frac{4\ell^2 (M_{b}-M_{n})}{r^7}
\nonumber \\
 &=&\frac{4 (M_{b}-M_{n})}{\ell^5}.
\end{eqnarray}
Using $\sqrt{g}=(r^2-n^2)^2 \sin{\theta_1}\sin{\theta_2}$, one has
\begin{equation}d\widetilde{S}_{\tau}= \ell^4
\sin{\theta_1}\sin{\theta_2}d\theta_1 d\theta_2 d\phi_1
d\phi_2 .
\end{equation}
Finally we obtain the mass of the six dimensional Bolt solution
\begin{eqnarray}
\Delta M &=&\frac{\ell}{8\pi G}\int_{\partial \Sigma}\frac{4
(M_{b}-M_{n})}{\ell^5}\ell^4 \sin{\theta_1}\sin{\theta_2}d\theta_1
d\theta_2 d\phi_1 d\phi_2 \nonumber \\
&=&\frac{8\pi}{G}(M_{b}-M_{n}).
\end{eqnarray}
Again, it is identical to the one in \cite{Mann1}.

\subsection{Eight Dimensional Solutions}
The eight dimensional Nut charged AdS solution has the following
form~\cite{AAwad}
\begin{eqnarray}
\label{tubnut8} ds^2 &=& F(r) (d \tau + 2n \cos{\theta_1}
d{\phi}_1 + 2n \cos{ \theta}_2d { \phi }_2+2n \cos{ \theta}_3d {
\phi }_3)^2+ F(r)^{-1} dr^2
\nonumber\\
 &&+ (r^2-n^2)(d{\theta_1}^2+\sin^{2} \theta_{1} d
{\phi_1}^2+d{\theta_2}^2+\sin^{2} \theta_{2} d
{\phi_2}^2+d{\theta_3}^2+\sin^{3} \theta_{3} d {\phi_3}^2),
\end{eqnarray}
where $F(r)$ is given by
\begin{equation}
F(r)={5r^8+(\ell^2-28n^2)r^6+5n^2(14n^2-\ell^2)r^4+5(3\ell^2-28n^2)r^2-10Mr\ell^2+5n^6(\ell^2-7
n^2)\over 5\ell^2(r^2-n^2)^3}.
\end{equation}
In order to have a Nut solution, the mass parameter $M$ must be fixed as
\begin{equation}
M_n={8n^5(\ell^2-8n^2) \over 5\ell^2}.
\end{equation}
Once again by fixing the mass at the above value, the function
$F(r)$ is then
\begin{equation}
F(r)={(r-n)(5r^4+20nr^3+(\ell^2+22n^2)r^2+(4n\ell^2-12n^3)r-35n^4+5\ell^2n^2)
\over 5(r+n)^3\ell^2}.
\end{equation}
In order to have a regular Bolt solution we must impose the mass
$M$ as
\begin{equation}
M_b={ 1 \over
10\ell^2}[{r_b}^7+(\ell^2-28n^2){r_b}^5+5n^2(14n^2-\ell^2){r_b}^3+5(3\ell^2-28n^2)r_b+5n^6(\ell
^2-7n^2)/r_b],
\end{equation}
where $r_b$ is
\begin{equation}
r_{b \pm}={ 1\over 56 n }\left(\ell^2\pm
\sqrt{\ell^4-448n^2\ell^2+3136n^4}\right).
\end{equation}
Requiring that $r_b$ be real and also be greater than $n$ implies that
\begin{equation}
n \le \left({4-\sqrt{15} \over 56} \right)^{1\over 2}\ell.
\end{equation}
Treating the Nut solution as a reference solution, we can get the
conserved charges for the Bolt solution as
\begin{equation}
H_{\partial_{\tau}}[g_1]=\frac{\ell}{8 \pi G \cdot 5}
\int_{\partial \Sigma}
\Omega^{-5}\left[(\widetilde{C}_{1})_{acbd}(\widetilde{n}_1)^{c}
(\widetilde{n}_1)^{d}-(\widetilde{C}_{0})_{acbd}(\widetilde{n}_0)^{c}
(\widetilde{n}_0)^{d}\right]\widetilde{u}^b (\partial_{\tau})^a \,
d \widetilde{S}.
\end{equation}
Note that in this case one has the following component of the Weyl
tensor for the solution (\ref{tubnut8})
\begin{eqnarray}
\widetilde{C}^{\tau}{}_{r\tau r}=C^{\tau}{}_{r\tau r}&\sim&
\frac{6/5n^2(\ell^2-8n^2)}{r^6}
\nonumber \\
 &&-\frac{3/175(14\ell^4n^2
+\ell^2(1875-1469n^4)+28n^2(-625+477n^4))}{r^8} \nonumber \\
&&+\frac{30 M \ell^2 }{r^9} +O(\frac{1}{r^{10}}),
\end{eqnarray}
and then
\begin{eqnarray}
(\widetilde{E}_{1})^{\tau}{}_{\tau}-(\widetilde{E}_0)^{\tau}{}_{\tau}&=&
\frac{1}{5}\Omega^{-5}\left[(C_{1})^{\tau}{}_{ r\tau r}
(\widetilde{n}_1)^{r}(\widetilde{n}_1)^{r}-(C_{0})^{\tau}{}_{
r\tau r}(\widetilde{n}_0)^{r}(\widetilde{n}_0)^{r}\right]  \nonumber \\
&\sim&\frac{r^9}{\ell^9}\frac{6\ell^2 (M_{b}-M_{n})}{r^9}
\nonumber \\
&=&\frac{6(M_{b}-M_{n})}{\ell^7}.
\end{eqnarray}
With $\sqrt{g}=(r^2-n^2)^3
\sin{\theta_1}\sin{\theta_2}\sin{\theta_3}$, and
\begin{equation}d\widetilde{S}_{\tau}= \ell^6
\sin{\theta_1}\sin{\theta_2}\sin{\theta_3}d\theta_1
d\theta_2d\theta_3 d\phi_1 d\phi_2d\phi_3,
\end{equation}
we obtain the energy of the eight dimensional Bolt solution
\begin{eqnarray}
\Delta M &=&\frac{\ell}{8\pi G}\int_{\partial \Sigma}\frac{6
(M_{b}-M_{n})}{\ell^7}\ell^6
\sin{\theta_1}\sin{\theta_2}\sin{\theta_3}d\theta_1
d\theta_2d\theta_3 d\phi_1 d\phi_2d\phi_3 \nonumber \\
&=&\frac{48\pi^2}{G}(M_{b}-M_{n}).
\end{eqnarray}

\subsection{Ten Dimensional Solutions}
Ten dimensional Nut charged AdS solution is given by~\cite{AAwad}
\begin{eqnarray}
ds^2 &=& F(r)(d\tau+2n\cos{\theta}_1 d{\phi}_1+2n\cos{\theta}_2
d{\phi}_2+2n\cos{\theta}_3d{\phi}_3+2n\cos{\theta}_4d{\phi}_4)^2\nonumber\\
&+&F(r)^{-1}dr^2+(r^2-n^2)(d{\theta_1}^2
+\sin^2{\theta}_1{d{\phi}_1}^2+{d{\theta}_2}^2+\sin^2{\theta}_2
{d{\phi}_2}^2 \nonumber \\
&+&{d{\theta}_3}^2+\sin^2{\theta}_3
{d{\phi}_3}^2+ {d{\theta}_4}^2+\sin^2{\theta}_4 {d{\phi}_4}^2),
\end{eqnarray}
where $F(r)$ has the form
\begin{eqnarray}
F(r)&=&{1\over
35\ell^2(r^2-n^2)^4}[35r^{10}+5(\ell^2-45n^2)r^8+14n^2(45n^2-2\ell^2)r^6
\nonumber \\
&+&70n^4(\ell^2-15n^2)r^4+35n^6(45n^2-4\ell^2)r^2-70Mr\ell^2+35n^8(9n^2-\ell^2)].
\end{eqnarray}
In order to describe a Nut solution, the mass parameter $M$ must be fixed as
\begin{equation}
M_n={64n^7(10n^2-\ell^2) \over 35\ell^2}.
\end{equation}
Taking the mass parameter to be the above value, one has $F(r)$ as
\begin{eqnarray}
F(r)&=&{1 \over
35(r+n)^4\ell^2}(r-n)(35r^5+175nr^4+(300n^2+5\ell^2)r^3
\nonumber \\
&+&(25n\ell^2+100n^3)r^2+(47n^2\ell^2-295n^4)r
-315n^5+35\ell^2n^3).
\end{eqnarray}
In order to have a regular Bolt solution we must impose the mass
$M$ as
\begin{eqnarray}
M_b&=&{ 1 \over 70\ell^2}[35{r_b}^9+(5\ell^2-225n^2){r_b}^7+
n^2(630n^2-28\ell^2){r_b}^5 +n^4(70\ell^2-1050n^2){r_b}^3\nonumber
\\ &+&n^6(1575n^2-140\ell^2)r_b+n^8(315n^2-35\ell^2)],
\end{eqnarray}
where
\begin{equation}
r_{b \pm}={ 1\over 90 n }\left(\ell^2\pm
\sqrt{\ell^4-900^2\ell^2+8100n^4}\right).
\end{equation}
Requiring that $r_b$ is real and larger than $n$ implies
\begin{equation}
n\le \left({ 5-2\sqrt{6}\over 90}\right)^{1\over2}\ell.
\end{equation}
In this case we have the conserved charge of the Bolt solution by
considering the Nut solution as a reference solution
\begin{equation}
H_{\partial_{\tau}}[g_1]=\frac{\ell}{8 \pi G \cdot 7}
\int_{\partial \Sigma}
\Omega^{-7}\left[(\widetilde{C}_{1})_{acbd}(\widetilde{n}_1)^{c}
(\widetilde{n}_1)^{d}-(\widetilde{C}_{0})_{acbd}(\widetilde{n}_0)^{c}
(\widetilde{n}_0)^{d}\right]\widetilde{u}^b (\partial_{\tau})^a \,
d \widetilde{S}.
\end{equation}
Note that here we have
\begin{eqnarray}
\widetilde{C}^{\tau}{}_{r\tau r}=C^{\tau}{}_{r\tau r}&\sim&
\frac{24n^2\left( \ell^2 - 10n^2 \right)} {35r^{6}} \nonumber \\
&+& \frac{8\left( -3\ell^4n^2 + 11\ell^2n^4 + 190n^6 \right)
}{245r^{8}} \nonumber \\
 &+& \frac{8\left( 15\ell^6n^2 -
142\ell^4n^4 + 22350\ell^2n^6 - 224300n^8 \right) }{8575r^{10}}
 \nonumber \\
&+& \frac{56\ell^2M}{r^{11}} + O(\frac{1}{r^{12}}),
\end{eqnarray}
and
\begin{eqnarray}
(\widetilde{E}_{1})^{\tau}{}_{\tau}-(\widetilde{E}_0)^{\tau}{}_{\tau}&=&
\frac{1}{7}\Omega^{-7}\left[(C_{1})^{\tau}{}_{ r\tau r}
(\widetilde{n}_1)^{r}(\widetilde{n}_1)^{r}-(C_{0})^{\tau}{}_{
r\tau r}(\widetilde{n}_0)^{r}(\widetilde{n}_0)^{r}\right]  \nonumber \\
&\sim&\frac{r^9}{\ell^9}\frac{8\ell^2 (M_{b}-M_{n})}{r^9}
\nonumber \\
&=&\frac{8(M_{b}-M_{n})}{\ell^7}.
\end{eqnarray}
In ten dimensional case, one has $\sqrt{g}=(r^2-n^2)^4
\sin{\theta_1}\sin{\theta_2}\sin{\theta_3}\sin{\theta_4}$, and
\begin{equation}d\widetilde{S}_{\tau}= \ell^8
\sin{\theta_1}\sin{\theta_2}\sin{\theta_3}\sin{\theta_4}d\theta_1
d\theta_2d\theta_3 d\theta_4 d\phi_1 d\phi_2 d\phi_3 d\phi_4.
\end{equation}
Thus, we obtain the mass of the ten dimensional Bolt solution
\begin{eqnarray}
\Delta M &=&\frac{\ell}{8\pi G}\int_{\partial \Sigma}\frac{8
(M_{b}-M_{n})}{\ell^9}\ell^8
\sin{\theta_1}\sin{\theta_2}\sin{\theta_3}d\theta_1
d\theta_2d\theta_3 d\phi_1 d\phi_2d\phi_3 \nonumber \\
&=&\frac{256\pi^3}{G}(M_{b}-M_{n}).
\end{eqnarray}

At the end of this section, we give the mass formula for a general
Nut charged AdS solution in $2k+2$ dimension, which has the
form~(\ref{tubnutgeneral}). The Nut solution is obtained by fixing
the mass parameter as~\cite{Mann1}
\begin{equation}
M_{n} =\frac{n^{2k-1}}{\sqrt{\pi }\ell ^{2}}\left[\ell
^{2}-(2k+2)n^{2}\right]\frac{ \Gamma \left( \frac{3-2k}{2}\right)
\Gamma \left( k+1\right) }{\left( 2k-1\right) },
\end{equation}
while the Bolt solution corresponds to the case with the mass
parameter
\begin{equation}
M_{b}=\frac{1}{2}\left[ \sum_{i=0}^{k}\left( {k \atop i } \right)
\frac{
(-1)^{i}n^{2i}r_{b}^{2k-2i-1}}{(2k-2i-1)}+\frac{(2k+1)}{\ell ^{2}}
\sum_{i=0}^{k+1}\left( {k+1 \atop i }\right) \frac{
(-1)^{i}n^{2i}r_{b}^{2k-2i+1}}{(2k-2i+1)}\right],
\end{equation}
where $r_{b}>n$ is determined by $F(r_{b})=0$ and $F^{\prime
}(r_{b})={\frac{2}{n(2k+2)}}$. For these general Taub-Bolt-AdS and
Taub-Nut-AdS solutions we find
\begin{equation}
\widetilde{E}^{\tau}{}_{1,\tau}-\widetilde{E}^{\tau}{}_{0,\tau}
=\frac{(n-2) (M_{b}-M_{n}) }{\ell^{n+1}},\ \, \quad
\Delta M=\frac{(n-2)(4\pi)^{(n-2)/2}(M_{b}-M_{n})}{8\pi G}.
\end{equation}
This result coincides with that in \cite{Mann1}, where those
authors get the result by using ``boundary counterterm" method and
``Noether method".

\section{Four Dimensional Kerr-Taub-Nut-AdS Solution}

In this section we discuss the case of  four dimensional Euclidean
Kerr-Taub-Nut-AdS space-times, which has the form~\cite{Mann}
\begin{eqnarray}
\label{eq115}
 ds^2 &=& \frac{V(r)\,({d{\tau }}
  -(2N\cos\theta-a\sin^2\theta)d\phi)^{2})
+ {\mathcal {H}}(\theta )\sin^2{\theta} \,(a\,{d{\tau }}
 - (r^{2} - N^{2} - a^{2})\,{d{\phi }})^{2}
}{\chi ^{4}\,(r^{2} - (N + a\,\cos{\theta })^{2})}\nonumber \\
&&+(r^{2} - (N + a\,\cos{\theta })^{2})\, \left( \frac
{dr^2}{V(r)}   + \frac {d\theta
^2}{ \mathcal {H}(\theta )} \right),
\end{eqnarray}
where
\begin{eqnarray}
\mathcal{H}(\theta) &=& 1 +  \frac {q\,N^{2}}{\ell^{2}} +
\frac{(2\,N + a\,\cos{\theta })^{2}}{\ell^{2}},   \nonumber \\
 V(r) &=&  \frac {r^{4}}{\ell^{2}}  +
\frac{((q - 2)N^{2} -a^2 +\ell^{2})\,r^{2}}{\ell^{2}} - 2\,M\,r -  \frac {(a
+ N)\,(a - N)\,(q\,N^{2} + \ell^{2} + N^{2})}{\ell^{2}}.
\end{eqnarray}
The periodicity in $\tau$ and the parameters $q$ and $\chi$ are
chosen so that conical singularities are avoided. In the
$(\theta,\phi)$ section these considerations imply that $q=-4$ and
$\chi=1/\sqrt{1+a^2/l^2}$.

For this solution, a straightforward calculation gives
\begin{eqnarray}
\widetilde{C}^{\tau}{}_{r \tau r}&=&C^{\tau}{}_{r \tau r}\sim
\frac{2 M \ell^2}{r^5}+O(\frac{1}{r^6}), \nonumber
\\
\widetilde{C}^{\tau}{}_{r \phi r}&=&C^{\tau}{}_{r \phi r}\sim
\frac{3 M \ell^2(a\sin^2{\theta} -2N
\cos{\theta})}{r^5}+O(\frac{1}{r^6}).
\end{eqnarray}
The conformal boundary volume has the form
\begin{equation}
d\widetilde{S}_{\tau}= \frac{\ell^2}{\chi^4} \sin{\theta}d\theta
d\phi.
\end{equation}
In this case we choose the massless solution, namely the case with
$M=0$, as a reference solution. Thus we find
\begin{eqnarray}
(\widetilde{E}_{1})^{\tau}{}_{\tau}-(\widetilde{E}_0)^{\tau}{}_{\tau}&=&
\Omega^{-1}\left[(C_{1})^{\tau}{}_{ r\tau r}
(\widetilde{n}_1)^{r}(\widetilde{n}_1)^{r}-(C_{0})^{\tau}{}_{
r\tau r}(\widetilde{n}_0)^{r}(\widetilde{n}_0)^{r}\right] \\
\nonumber &\sim&\frac{r^5}{\ell^5}\frac{2\ell^2 M}{r^5}=\frac{2
M}{\ell^3},
\end{eqnarray}
and the conserved charge associated to $\partial_\tau$
\begin{equation} H_{\partial_{\tau}} =\frac{\ell}{8\pi
G }\int_{\partial \Sigma}\frac{2 M}{\ell^3}\frac{\ell^2}{\chi^4}
\sin{\theta}d\theta d\phi= \frac{1}{G}\frac{M}{\chi^4}.
\end{equation}
Similarly we have
\begin{eqnarray}
(\widetilde{E}_{1})^{\tau}{}_{\phi}-(\widetilde{E}_0)^{\tau}{}_{\phi}&=&
\Omega^{-1}\left[(C_{1})^{\tau}{}_{ r\phi r}
(\widetilde{n}_1)^{r}(\widetilde{n}_1)^{r}-(C_{0})^{\tau}{}_{
r\phi r}(\widetilde{n}_0)^{r}(\widetilde{n}_0)^{r}\right]
\nonumber
\\
 &\sim&\frac{r^5}{\ell^5}\frac{3M( a
\sin^2{\theta}-2N\cos^2{\theta})\ell^2}{ \ r^5}=\frac{3 M ( a
\sin^2 {\theta}-2N\cos^2{\theta})}{ \ell^3},
\end{eqnarray}

\begin{equation}
H_{\partial_{\phi}} =\frac{\ell}{8\pi G}\int_{\partial \Sigma}
\frac{3 M (a \sin^2 {\theta}-2N\cos^2{\theta})}{\ell^3}
\frac{\ell^2}{\chi^4}\sin{\theta} d\theta d\phi =
\frac{1}{G}\frac{M a}{ \chi^4}.
\end{equation}
These results are just what Mann obtained in \cite{Mann}, but he
used the ``boundary counterterm" method.

It should be noted here, that the energy $E$ and angular momentum
$J_{\phi}$ are defined as $E=H_{\partial_{t}}$ and
$J_{\phi}=-H_{\partial_{\phi}}$ in the Lorentz sector. The
relative sign difference between definitions of energy and angular
momentum can be traced back to its origin for the Lorentz
signature of the space-time metric as mentioned in~\cite{Iwald}.
This can be understood by noting the definitions of energy and
angular momentum of a particle in special relativity: $E=-p_a t^a$
and $J=+p_a \phi^a$. In the Euclidean sector, the relative sign
difference disappears, and the Hamilton associated to
$\partial_{\phi}$ is just the angular momentum. In the next
section, we will calculate the energy and angular momentum of
higher dimensional Kerr-AdS solutions with Nut charges, and the
relative sign difference will appear because we consider the
solutions in the Lorentz sector.

\section{Higher Dimensional Kerr-AdS Solutions with Nut Charges}

The higher dimensional Kerr-AdS solution with Nut charges has been
given recently in~\cite{ZWChong}
\begin{equation}
ds^2 = \frac{p^2 + q^2}{X} dp^2 + \frac{p^2 + q^2}{Y} dq^2 +
\frac{X}{p^2 + q^2}(d\tau + q^2d\sigma)^2 - \frac{Y}{p^2 + q^2}
(d\tau -p^2 d\sigma)^2 + \frac{p^2 q^2}{\gamma} d\Omega_k^2,
\label{ktna}
\end{equation}
where \begin{equation}
 X=\gamma -\epsilon p^2 +\frac{1}{\ell^2}
p^4 + 2 N p^{1-k}\,,\qquad Y=\gamma +\epsilon q^2 +
\frac{1}{\ell^2} q^4 - 2 m\, q^{1-k},
\label{xy}
\end{equation}
and $d\Omega_k^2$ is the metric on the unit sphere $S^k$,
$(\gamma, m, N)$ are three independent continuous parameters,
which are related to the angular momentum, mass and Nut charge,
respectively, $\epsilon$ is a dimensionless constant, and $\ell$
is the AdS radius. If we take the parameters in (\ref{xy}) to be
\begin{equation}
\gamma=a^2\,,\qquad \epsilon= 1+ \frac{1}{\ell^2} a^2 \,,\qquad
m=M,
\end{equation}
 define
\begin{eqnarray} \Delta_r
&=&(r^2+a^2)(1+\frac{r^2}{\ell^2})-2M r^{1-k}, \nonumber \\
\Delta_{\theta}&=&1-\frac{a^2}{\ell^2}\cos^2{\theta}, \nonumber
\\
\Xi &=& 1-\frac{a^2}{\ell^2}, \nonumber  \\
\rho^2 &=&r^2+a^2\cos^2{\theta},
\end{eqnarray}
and choose a set of new coordinates as
\begin{equation}
\label{coordtrans} p=a\,\cos\theta\,,\qquad q=r\,,\qquad \tau = t
-\frac{a}{\Xi}\, \phi\,,\qquad \sigma = - \frac{1}{a\, \Xi}\,
\phi\,,
\end{equation}
then we can rewrite the solution (\ref{ktna}) in the form
\begin{eqnarray}
\label{eq128}
ds^2&=&-\frac{\Delta_{r}}{\rho^2}\left(dt-\frac{a}{\Xi}\sin^2{\theta}d\phi\right)^2
+\frac{\rho^2}{\Delta_{r}}dr^2+\frac{\rho^2}{\Delta_{\theta}+\frac{2N(a
\cos{\theta})^{1-k}}{a^2\sin^2{\theta}}}d\theta^2 \nonumber \\
&+&
\frac{\Delta_{\theta}\sin^2{\theta}+\frac{2N(a\cos{\theta})^{1-k}}{a^2}}{\rho^2}
\left(a
dt-\frac{a^2+r^2}{\Xi}d\phi\right)^2+r^2\cos^2{\theta}d\Omega^2_{k}.
\end{eqnarray}
When $N=0$, this metric reduces to the metric obtained in
\cite{hawkinghunter}, which describes a higher dimensional
Kerr-AdS black hole with a single rotation parameter. Note that in
the four dimensional case, this solution (\ref{eq128}) seems not
completely the same as that given in (\ref{eq115}).

Taking the massless Kerr-Taub-Nut-AdS as the reference solution,
namely the solution with $M=0$, we can calculate the conserved
quantities associated to killing vector fields $\partial_{t}$ and
$\partial_{\phi}$ of the solution (\ref{ktna}) in the coordinates
$(t , r , \theta ,\phi ,\cdots )$, according to the following
formulas
\begin{equation}
H_{\partial_{t}}[g_1]=\frac{\ell}{8 \pi G (n-3)} \int_{\partial
\Sigma} \lim_{\rightarrow
\mathcal{B}}\Omega^{3-n}\left[(\widetilde{C}_{1})_{acbd}(\widetilde{n}_1)^{c}
(\widetilde{n}_1)^{d}-(\widetilde{C}_{0})_{acbd}(\widetilde{n}_0)^{c}
(\widetilde{n}_0)^{d}\right]\widetilde{u}^b (\partial_{t})^a \, d
\widetilde{S},
\end{equation}
and
\begin{equation}
H_{\partial_{\phi}}[g_1]=\frac{\ell}{8 \pi G (n-3)} \int_{\partial
\Sigma}\lim_{\rightarrow \mathcal{B}}
\Omega^{3-n}\left[(\widetilde{C}_{1})_{acbd}(\widetilde{n}_1)^{c}
(\widetilde{n}_1)^{d}-(\widetilde{C}_{0})_{acbd}(\widetilde{n}_0)^{c}
(\widetilde{n}_0)^{d}\right]\widetilde{u}^b (\partial_{\phi})^a \,
d \widetilde{S},
\end{equation}
where the conformal factor is taken to be $\frac{\ell}{r}$.

\subsection{Four Dimensional  Solutions}
In four dimensional case, we find
\begin{eqnarray}
\widetilde{C}^{t}{}_{r t r}&=&\frac{2 M
\ell^2}{r^5}+O(\frac{1}{r^6}), \nonumber \\
\widetilde{C}^{t}{}_{r\phi r}&=&\frac{ -3 M
a\sin^2{\theta}\ell^2}{\Xi r^5}+O(\frac{1}{r^6}).
\end{eqnarray}
The conformal boundary volume has the form
\begin{equation}
d\widetilde{S}_{t}= \frac{\ell^2}{\Xi} \sin{\theta}d\theta d\phi,
\end{equation}
and
\begin{eqnarray}
(\widetilde{E}_{1})^{t}{}_{t}-(\widetilde{E}_0)^{t}{}_{t}&=&
\Omega^{-1}\left[(C_1)^{t}{}_{r t r} (\widetilde{n}_1)^{r}
(\widetilde{n}_1)^{r}-(C_0)^{t}{}_{ r t r}
(\widetilde{n}_0)^{r}(\widetilde{n}_0)^{r}\right], \nonumber
\\
&\sim&\frac{r^5}{\ell^5}\frac{2\ell^2 M}{r^5}=\frac{2 M}{\ell^3}.
\end{eqnarray}
Thus the conserved charge associated to $\partial_t$ is
\begin{equation} H_{\partial_{t}} =\frac{\ell}{8\pi
G}\int_{\partial \Sigma}\frac{2 M}{\ell^3}\frac{\ell^2}{\Xi}
\sin{\theta}d\theta d\phi= \frac{1}{G}\frac{M}{\Xi}.
\end{equation}
Similarly we have
\begin{eqnarray}
(\widetilde{E}_1)^{t}{}_{\phi}-(\widetilde{E}_0)^{t}{}_{\phi}&=&
\Omega^{-1}\left[(C_1)^{t}{}_{r\phi r}
(\widetilde{n}_1)^{r}(\widetilde{n}_1)^{r}-(C_0)^{t}{}_{ r\phi r}
(\widetilde{n}_0)^{r}(\widetilde{n}_0)^{r}\right]  \nonumber
\\
 &\sim&-\frac{r^5}{\ell^5}\frac{-3 M a \sin^2{\theta} \ell^2 }{\Xi
r^5}=-\frac{3 M a \sin^2{\theta}}{\Xi \ell^3},
\end{eqnarray}
and
\begin{equation}
H_{\partial_{\phi}} =\frac{\ell}{8\pi G}\int_{\partial \Sigma}
\frac{3 m \sin^2{\theta}}{\Xi\ell^3}
\frac{\ell^2}{\Xi}\sin{\theta} d\theta d\phi= -\frac{1}{G}\frac{M
a}{\Xi^2}.
\end{equation}
Thus the associated angular momentum is
\begin{equation}
J_{\phi} =-H_{\partial_{\phi}}=\frac{1}{G}\frac{Ma}{\Xi^2}.
\end{equation}
According to the definition~\cite{Gibbons}, the ``conformal mass"
of the solution is
\begin{equation}
M_{c}=H_{\partial_{t}}+\frac{a}{\ell^2}J_{\phi}=\frac{1}{G}\frac{M}{\Xi^2},
\end{equation}
which satisfies the first law of thermodynamics. We note that the
mass and angular momentum of the Kerr-Taub-Nut-AdS solution have
completely the same form as those of the Kerr-AdS solution.

\subsection{Six Dimensional Solutions}

For six dimensional solution, we have
\begin{eqnarray}
\widetilde{C}^{t}{}_{r t r}&=&\frac{6N \ell^2}{r^6 a\cos{\theta}
}+\frac{12 M \ell^2}{r^7}+O(\frac{1}{r^8}), \nonumber \\
\widetilde{C}^{t}{}_{r \phi r}&=& -\frac{7N\ell^2
\sin^2{\theta}}{\Xi r^6\cos{\theta}}-\frac{15 M a
\sin^2{\theta}}{\Xi r^7}+O(\frac{1}{r^8}).
\end{eqnarray}
The conformal boundary volume takes the form
\begin{equation}
d\widetilde{S}_{t}= \frac{\ell^4}{\Xi} \cos^2{\theta} \sin{\theta}
d\theta d\phi d\Omega_{2},
\end{equation}
and
\begin{eqnarray}
(\widetilde{E}_{1})^{t}{}_{t}-(\widetilde{E}_0)^{t}{}_{t}&=&
\frac{1}{3}\Omega^{-3}\left[(C_1)^{t}{}_{r t r}
(\widetilde{n}_1)^{r} (\widetilde{n}_1)^{r}-(C_0)^{t}{}_{ r t r}
(\widetilde{n}_0)^{r}(\widetilde{n}_0)^{r}\right]  \nonumber \\
&\sim&\frac{r^7}{\ell^7}\frac{4\ell^2 M}{r^7}=\frac{4 M}{\ell^5}.
\end{eqnarray}
Thus we obtain the conserved charge associated to $\partial_t$
\begin{equation} H_{\partial_{t}} =\frac{\ell}{8\pi
G }\int_{\partial \Sigma}\frac{4 M}{\ell^5}\frac{\ell^4}{\Xi}
\cos^2 {\theta}\sin{\theta}d\theta d\phi d\Omega_{2}=
\frac{1}{G}\frac{8\pi M}{3\Xi}.
\end{equation}
Similarly we find
\begin{eqnarray}
(\widetilde{E}_1)^{t}{}_{\phi}-(\widetilde{E}_0)^{t}{}_{\phi}&=&
\frac{1}{3}\Omega^{-3}\left[(C_1)^{t}{}_{r\phi r}
(\widetilde{n}_1)^{r}(\widetilde{n}_1)^{r}-(C_0)^{t}{}_{ r\phi r}
(\widetilde{n}_0)^{r}(\widetilde{n}_0)^{r}\right]  \nonumber \\
&\sim&-\frac{r^7}{\ell^7}\frac{5M a \sin^2{\theta}\ell^2}{\Xi \ \
r^7}=-\frac{5 M a \sin^2 {\theta}}{\Xi \ell^5},
\end{eqnarray}
\begin{equation}
H_{\partial_{\phi}} =\frac{\ell}{8\pi G}\int_{\partial \Sigma}
\frac{5 M a \sin^2 {\theta}}{\Xi\ell^5} \frac{\ell^4}{\Xi}\cos^2
{\theta}\sin{\theta} d\theta d\phi d\Omega_{2}= -\frac{1}{G}\frac{4
\pi M a}{3 \Xi^2}.
\end{equation}
The associated  angular momentum is then given by
\begin{equation}
J_{\phi} =-H_{\partial_{\phi}}=\frac{1}{G}\frac{4\pi M a}{3 \Xi^2}.
\end{equation}
and the ``conformal mass"  of the solution is
\begin{equation}
M_{c}=H_{\partial_{t}}+\frac{a}{\ell^2}J_{\phi}=\frac{4\pi M}{3 G \Xi}(1+\frac{1}{\Xi}).
\end{equation}

\subsection{Eight Dimensional Solutions}

In this case we have
\begin{eqnarray}
\widetilde{C}^{t}{}_{r t r}&=&\frac{6N \ell^2}{r^6 a^3
\cos^3{\theta} }-\frac{N \ell^2(a^2+6\ell^2
+19a^2\cos^2{\theta})}{r^8 a^3\cos^3{\theta}}+\frac{30 M
\ell^2}{r^9}+O(\frac{1}{r^{10}}),
\nonumber \\
\widetilde{C}^{t}{}_{r \phi r}&=& -\frac{5 N \ell^2
\sin^2{\theta}}{ r^6 a^2\cos^3{\theta}
\Xi}+\frac{N\ell^2(5\ell^2+22 a^2 \cos^2{\theta})}{ r^8a^2
\cos^3{\theta} \Xi}-\frac{35 M a \ell^2 \sin^2{\theta}}{ r^9
\Xi}+O(\frac{1}{r^{10}}),
\end{eqnarray}
and
\begin{eqnarray}
(\widetilde{E}_{1})^{t}{}_{t}-(\widetilde{E}_0)^{t}{}_{t}&=&
\frac{1}{5}\Omega^{-5}\left[(C_1)^{t}{}_{r t r}
(\widetilde{n}_1)^{r} (\widetilde{n}_1)^{r}-(C_0)^{t}{}_{ r t r}
(\widetilde{n}_0)^{r}(\widetilde{n}_0)^{r}\right] \\ \nonumber
&\sim&\frac{r^9}{\ell^9}\frac{6\ell^2 M}{r^9}=\frac{6 M}{\ell^7}.
\end{eqnarray}
Note that the conformal boundary volume has the form
\begin{equation}
d\widetilde{S}_{t}= \frac{\ell^6}{\Xi} \cos^4{\theta} \sin{\theta}
d\theta d\phi d\Omega_{4}.
\end{equation}
We obtain the conserved charge associated to $\partial_t$
\begin{equation} H_{\partial_{t}} =\frac{\ell}{8\pi
G }\int_{\partial \Sigma}\frac{6 M}{\ell^7}\frac{\ell^6}{\Xi}
\cos^4 {\theta}\sin{\theta}d\theta d\phi d\Omega_{4}=
\frac{1}{G}\frac{8\pi^2 M}{5\Xi}.
\end{equation}
Similarly we find
\begin{eqnarray}
(\widetilde{E}_1)^{t}{}_{\phi}-(\widetilde{E}_0)^{t}{}_{\phi}&=&
\frac{1}{5}\Omega^{-5}\left[(C_1)^{t}{}_{r\phi r}
(\widetilde{n}_1)^{r}(\widetilde{n}_1)^{r}-(C_0)^{t}{}_{ r\phi r}
(\widetilde{n}_0)^{r}(\widetilde{n}_0)^{r}\right] \\ \nonumber
&\sim&-\frac{r^9}{\ell^9}\frac{7 M a \sin^2{\theta}\ell^2}{\Xi \ \
r^9}=-\frac{7 M a \sin^2 {\theta}}{\Xi \ell^7},
\end{eqnarray}
\begin{equation}
H_{\partial_{\phi}} =\frac{\ell}{8\pi G}\int_{\partial \Sigma}
\frac{7 M a \sin^2 {\theta}}{\Xi\ell^7} \frac{\ell^6}{\Xi}\cos^4
{\theta}\sin^3{\theta} d\theta d\phi d\Omega_{4}=
-\frac{1}{G}\frac{8 \pi^2 M a}{15 \Xi^2}.
\end{equation}
Thus the angular momentum is
\begin{equation}
J_{\phi} =-H_{\partial_{\phi}}=\frac{1}{G}\frac{8 \pi^2 M a}{15
\Xi^2},
\end{equation}
and the  ``conformal mass" of the solution
\begin{equation}
M_{c}=H_{\partial_{t}}+\frac{a}{\ell^2}J_{\phi}=\frac{8\pi^2 M}{15 G \Xi}(2+\frac{1}{\Xi}).
\end{equation}

\subsection{Ten Dimensional Solutions}
This case gives
\begin{eqnarray}
\widetilde{C}^{t}{}_{r t r}&=& \frac{6N\ell^2}{r^6
a^5{\cos^5{\theta}}} - \frac{N\ell^2 \left( 6 \ell^2 +3 a^2+
17a^2\cos^2{\theta} \right) } {r^8 a^5\cos ^5\theta},
\nonumber \\
&+&  \frac{N\ell^2 \left(3\left( a^4 + a^2 \ell^2 +
2\ell^4 \right)  -  a^2\left( a^2 - 17\ell^2 \right) \cos^2
{\theta} +
  40a^4\cos^4{\theta}\right)} {r^{10}a^5{\cos^5{\theta}}}
\nonumber \\
&+& \frac{56 M \ell^2}{r^{11}} + O(\frac{1}{r^{12}})
\nonumber \\
\widetilde{C}^{t}{}_{r \phi r}&=& -\frac{3 N \ell^2
\sin^2{\theta}}{ r^6 a^4\cos^5{\theta} \Xi}+\frac{3
N\ell^2(\ell^2+6 a^2 \cos^2{\theta})\sin^2{\theta}}{ r^8a^4
\cos^5{\theta} \Xi}
\nonumber \\
&-&\frac{3 N\ell^2(\ell^4+6 a^2\ell^2
\cos^2{\theta}+15a^2 \cos^4{\theta})\sin^2{\theta}}{ r^{10}a^4
\cos^5{\theta} \Xi}-\frac{63 M a \ell^2 \sin^2{\theta}}{ r^{11}
\Xi}+O(\frac{1}{r^{12}}).
\end{eqnarray}
The conformal boundary volume has the form
\begin{equation}
d\widetilde{S}_{t}= \frac{\ell^8}{\Xi} \cos^6{\theta} \sin{\theta}
d\theta d\phi d\Omega_{6},
\end{equation}
Note that
\begin{eqnarray}
(\widetilde{E}_{1})^{t}{}_{t}-(\widetilde{E}_0)^{t}{}_{t}&=&
\frac{1}{7}\Omega^{-7}\left[(C_1)^{t}{}_{r t r}
(\widetilde{n}_1)^{r} (\widetilde{n}_1)^{r}-(C_0)^{t}{}_{ r t r}
(\widetilde{n}_0)^{r}(\widetilde{n}_0)^{r}\right]  \nonumber \\
&\sim&\frac{r^{11}}{\ell^{11}}\frac{8\ell^2 M}{r^{11}}=\frac{8
M}{\ell^9}.
\end{eqnarray}
We find the conserved charge associated to the Killing vector
$\partial_t$
\begin{equation} H_{\partial_{t}} =\frac{\ell}{8\pi
G }\int_{\partial \Sigma}\frac{8 M}{\ell^9}\frac{\ell^8}{\Xi}
\cos^6 {\theta}\sin{\theta}d\theta d\phi d\Omega_{6}=
\frac{1}{G}\frac{64\pi^3 M}{105\Xi}.
\end{equation}
Similarly one can get
\begin{eqnarray}
(\widetilde{E}_1)^{t}{}_{\phi}-(\widetilde{E}_0)^{t}{}_{\phi}&=&
\frac{1}{7}\Omega^{-7}\left[(C_1)^{t}{}_{r\phi r}
(\widetilde{n}_1)^{r}(\widetilde{n}_1)^{r}-(C_0)^{t}{}_{ r\phi r}
(\widetilde{n}_0)^{r}(\widetilde{n}_0)^{r}\right] \nonumber \\
&\sim&-\frac{r^{11}}{\ell^{11}}\frac{9 M a
\sin^2{\theta}\ell^2}{\Xi \ \ r^{11}}=-\frac{9 M a \sin^2
{\theta}}{\Xi \ell^9},
\end{eqnarray}
\begin{equation}
H_{\partial_{\phi}} =\frac{\ell}{8\pi G}\int_{\partial \Sigma}
\frac{9 M a \sin^2 {\theta}}{\Xi\ell^9} \frac{\ell^8}{\Xi}\cos^6
{\theta}\sin^3{\theta} d\theta d\phi d\Omega_{6}=-
\frac{1}{G}\frac{16 \pi^3 M a}{105 \Xi^2}.
\end{equation}
Thus the solution has the angular momentum
\begin{equation}
J_{\phi} =-H_{\partial_{\phi}}=\frac{1}{G}\frac{16 \pi^3 M a}{105
\Xi^2},
\end{equation}
and the ``conformal mass"
\begin{equation}
M_{c}=H_{\partial_{t}}+\frac{a}{\ell^2}J_{\phi}=\frac{16\pi^3 M}{105 G \Xi}(3+\frac{1}{\Xi}).
\end{equation}

\section{(Un)Wrapped Brane}
The black brane solutions with flat transverse space in
$n$-dimensions are also AlAdS space-times. Their metrics can be
written as~\cite{Birm,Horowitz}
\begin{equation}
ds^2=-\Delta(r)^2dt^2+\frac{dr^2}{\Delta(r)^2}+r^2({dx_1}^2 +\cdots
+{dx_{n-2}}^2)
\end{equation}
where $\Delta(r)^2=-2M/r^{n-3}+r^2/\ell^2$. In this metric, at
least one of the transverse direction $x_i$ should be compactified
so that parameter $M$ cannot be changed by rescaling the
coordinates. Thus, the non-vanishing conserved charge is the mass
only. Choosing the background reference solution as $M=0$,  we
have
\begin{equation}
\widetilde{E}^{t}{}_{t} \sim \frac{(n-2)M}{\ell^{n-1}}.
\end{equation}
The boundary conformal volume element has the form
\begin{equation}
d\widetilde{S}_t=\ell^{n-2} dv,
\end{equation}
where $dv$ denotes the volume element of $n-2$-transverse space.
We thus obtain
\begin{equation}
H_{\partial _{t}}=\frac{\ell}{8\pi G}\int \ \
\frac{(n-2)M}{\ell^{n-1}} \ell^{n-2} dv=\frac{(n-2)M}{8\pi G}V,
\end{equation}
where $V$ represents the volume of $n-2$-transverse space. This
result is the same as that of~\cite{Aros, Aros1}. Note that this
solution gives an example which has a Ricci flat boundary, so that
the Fefferman-Graham expansion without log-terms can be done in
any dimension~\cite{Galloway}, and the conserved charges can be
defined as in (\ref{cccon}).

\section{Conclusion and Discussion}
In this paper, based on the work of Hollands {\it et
al.}~\cite{Hollands}, we derive a formula of calculating conserved
charges in even dimensional asymptotically {\it locally} Anti-de
Sitter (AlAdS) space-times by using the covariant phase space
definition of Wald and Zoupas~\cite{Iwald}. Our formula
generalizes the formula proposed by Ashtekar {\it et
al.}~\cite{am, AD}. This formula is background dependent. We
therefore have to specify a reference solution when we calculate
conserved charges for a certain AlAdS space-time. Using this
formula we calculate the masses of Taub-Bolt-AdS space-times by
treating Taub-Nut-AdS space-times as reference solutions. The
resulting masses agree with those obtained previously by
``background subtraction" method and ``boundary counterterm"
method. We also discussed the conserved charges in four
dimensional Kerr-Taub-Nut-AdS solutions and  higher dimensional
Kerr-AdS solutions with Nut charges by treating the corresponding
massless solutions as the background reference solutions. For
these higher dimensional Kerr-AdS solutions with Nut charges,
these conserved charges are obtained at the first time. In
addition, as a further example of AlAdS space-times, the mass of
the (un)wrapped brane solutions in any dimensions is also studied.

It is interesting to discuss the odd dimensional case. However,
some log terms will appear in the Fefferman-Graham expansion in
this case. Therefore it will fail by naively applying the same
procedure as is exhibited in this paper to the odd dimensional
case. Nevertheless, we have found that the similar conserved
charges can be defined for any dimension if the static AlAdS
space-times have Ricci flat boundaries.

Topological AdS black holes also belong to a kind of AlAdS
space-times, and they may have nontrivial horizons and boundary
topologies (see for example \cite{Amin,Lemos,Cai,topo,Birm}). This
study is motivated by the discovery of
Ba\~nados-Teitelboim-Zanelli (BTZ) black holes \cite{BTZ}, which
are exact solutions in the three-dimensional Einstein gravity with
a negative cosmological constant, and are locally equivalent to a
three-dimensional anti-de Sitter space. The method of Ashtekar
{\it et al.} can not be used directly to compute the conserved
charges of these black hole space-times. So, it is interesting to
discuss the conserved charges of these solutions by using our new
formula.

The examples discussed in this paper suggest that in even
dimensional AlAdS space-times, a relation  of ``surface
counterterm method" to our method
\begin{equation}
H_{\xi}[g]=Q_{\xi}[g]-Q_{\xi}[g_{0}],
\end{equation}
where $Q_{\xi}[g]$ is the conserved charge obtained by using the
``surface conterterm" method for a solution $g$ and $Q_{\xi}[g_0]$
is the one corresponding to the selected reference solution
$g_{0}$. If $g$ is an AAdS space-time, this relation reduces to
\begin{equation}
H_{\xi}[g]=Q_{\xi}[g]-Q_{\xi}[AdS].
\end{equation}
This case has been discussed by  Hollands,  Ishibashi and Marolf
in the paper \cite{Hollands1}. They have used general arguments
based on the Peierls bracket to compare the counterterm charges
and the Hamiltonian charges defined in \cite{Hollands} in any
dimensional AAdS space-time. In the even dimensional AAdS case,
the counterterm charge of exact AdS space-times vanishes, while in
odd dimensional case the counterterm charge $Q_{\xi}[AdS]$
corresponds to the Casimir energy of the boundary CFTs. So, it is
interesting to find to what the counterterm charges
$Q_{\xi}[g_{0}]$ correspond in the boundary CFTs as one considers
the general AlAdS cases.

\section*{Acknowledgements}
L.M.Cao thanks Hong-Sheng Zhang, Hao Wei, Hui Li, Da-Wei Pang and
Yi Zhang for useful discussions and kind help. This work is
supported by grants from NSFC, China (No. 10325525 and No.
90403029), and a grant from the Chinese Academy of Sciences.

\end{document}